\documentclass[useAMS,usenatbib]{mn2e}

\usepackage{psfig,epsfig}
\usepackage{mathrsfs}
\usepackage{amssymb}
\usepackage{epsfig}
\usepackage{marvosym}
\usepackage{subfigure}
\usepackage{fancyhdr}
\usepackage{rotating}

\def\degr{\hbox{$^\circ$}}

\def\farcs{\hbox{$.\!\!^{\prime\prime}$}}
\def\gsim{\mathrel{\hbox{\rlap{\lower.55ex \hbox {$\sim$}}
                   \kern-.3em \raise.4ex \hbox{$>$}}}}
\def\lsim{\mathrel{\hbox{\rlap{\lower.55ex \hbox {$\sim$}}
                   \kern-.3em \raise.4ex \hbox{$<$}}}}

\def\msun{\hbox{M$_{\odot}\,$}}
\def\he{\hbox{He\,{\sc i} $\lambda$5875}}

\def\UVEX{\hbox{\sl UVEX\ }}
\def\IPHAS{\hbox{\sl IPHAS\ }}
\def\RL1{\hbox{{$R_{L_{1}}$}}}

\title[A first catalogue of automatically selected UV-excess sources from the UVEX survey]
{A first catalogue of automatically selected UV-excess sources from the UVEX survey}
\author[Kars Verbeek et al.]{{Kars Verbeek$^{1}$\thanks{E-mail:k.verbeek@astro.ru.nl},
Eelco de Groot$^{1}$,
Paul J. Groot$^{1}$, 
Simone Scaringi$^{1}$,
Janet Drew$^{2}$,}
\newauthor{
Robert Greimel$^{3}$,
Mike Irwin$^{4}$,
Eduardo Gonz{\'a}lez-Solares$^{4}$,
Boris T. G\"ansicke$^{5}$,}
\newauthor{
Jorge Casares$^{6}$, 
Jesus M. Corral-Santana$^{6,7}$,
Niall Deacon$^{8}$
and Danny Steeghs$^{5}$}\\
$^{1}$Department of Astrophysics, Radboud University Nijmegen,
  P.O. Box 9010, 6500 GL Nijmegen, The Netherlands\\
$^{2}$Centre for Astronomy Research, Science \& Technology Research
  Institute, University of Hertfordshire, Hatfield, AL10 9AB, UK\\
$^{3}$Institut f\"ur Physik, Karl-Franzen Universit\"at Graz,
Universit\"atsplatz 5, 8010 Graz, Austria\\
$^{4}$Cambridge Astronomy Survey Unit, Institute of Astronomy, University of
  Cambridge, Madingley Road, Cambridge, CB3 0HA, UK\\
$^{5}$Physics Department, University of Warwick, Coventry, CV4 7AL,
  UK\\
$^{6}$Instituto de Astrof\'{\i}sica de Canarias, Via Lactea, s/n
  E-38205 La Laguna (Tenerife), Spain\\
$^{7}$Departamento de Astrof\'{\i}sica, Universidad de La Laguna, 
  La Laguna E-38205, S/C de Tenerife, Spain\\
$^{8}$Institute for Astronomy, University of Hawaii, 
  2680 Woodlawn Drive, Honolulu, HI 96822, USA
}

\begin{document}

\date{Accepted ...  Received ...; in original form ...}

\pagerange{\pageref{firstpage}--\pageref{lastpage}} \pubyear{2011}

\maketitle

\label{firstpage}

\begin{abstract}
We present the first catalogue of point-source UV-excess sources selected from the \UVEX survey.
\UVEX images the Northern Galactic Plane in the $U,g,r$ and $\he$ bands in the Galactic
latitude range --5\degr$<$ $b$ $<$+5\degr. Through an automated algorithm, which works on a
field-to-field basis, we select blue UV-excess sources in 211 square degrees from the $(U-g)$ vs. $(g-r)$
colour-colour diagram and the $g$ vs. $(U-g)$ and $g$ vs. $(g-r)$ colour-magnitude diagrams. 
The UV-excess catalogue covers the magnitude range $14 < g < 22.5$, contains 2\,170 sources and
consists of a mix of white dwarfs, post-common-envelope objects, interacting binaries, quasars and AGN.
Two other samples of outliers were found during the selection: 
i) a `subdwarf' sample, consisting of 
no less than 9\,872 candidate metal-poor stars or lightly reddened main-sequence stars,
and ii) a `purple' sample consisting of 803 objects, most likely a mix of 
reddened late M-giants, T Tauri stars, planetary nebulae, symbiotic stars and carbon stars. 
Cross-matching the selected UV-excess catalogue with other catalogues 
aids with the first classification of the different populations 
and shows that more than $99\%$ of our selected sources are unidentified sources.
\end{abstract}


\section{Introduction}
The UV-excess survey of the Northern Galactic Plane (\UVEX)\footnote{http://www.uvexsurvey.org}
images a strip of 10$\times$185 degrees (--5\degr$<$ $b$ $<$+5\degr) along the Northern Galactic plane in the $U,g,r$ and $\he$
bands down to $\sim 21^{st}-22^{nd}$ magnitude using the Wide Field Camera mounted on the Isaac Newton Telescope, on the island of La Palma.
The spatial pixel scale is 0.33 arcsec/pixel with a field of view of 0.29 square degree per pointing.
The exposure times are 30 seconds for the $r$- and $g$-band images, 120 seconds 
for the $U$-band images and 180 seconds for the $\he$-band images.
A full description of the survey is given in Groot et al. (2009; hereafter G09).\\

\begin{figure*}
\centerline{\psfig{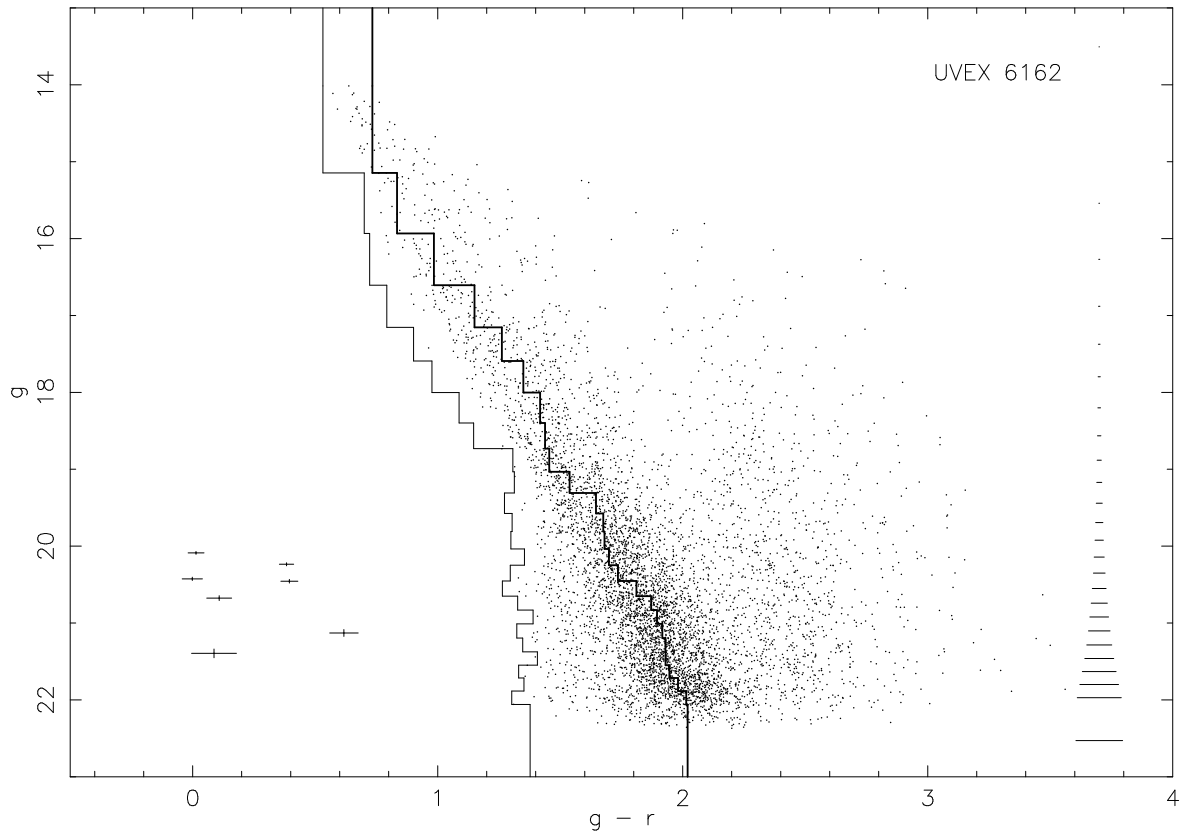}}
\caption{$g$ vs. $(g-r)$ colour-magnitude diagram of field UVEX 6162 
($l$=80.76, $b$ = --1.90). Data shown are all stellar and probably stellar 
detections in the $r$- and $g$-bands with a photometric error smaller than 0.1 magnitudes in these bands. 
The solid lines show our automated selection technique and the 
outline of the $(g-r)$ colour at which the $g$-band magnitude peaks (right, thick 
line), and the 3$\sigma$ limit for each $g$-band magnitude bin (left thin line),
called the `blue edge'. All sources more than 3 times their own photometric error to 
the left of this blue edge are selected (here plotted with 
their own photometric errors). The error bars on the right side show the 
average photometric error in $(g-r)$ per bin.  
\label{fig:uvex6162cmd}}
\end{figure*}

\begin{figure*}
\centerline{\psfig{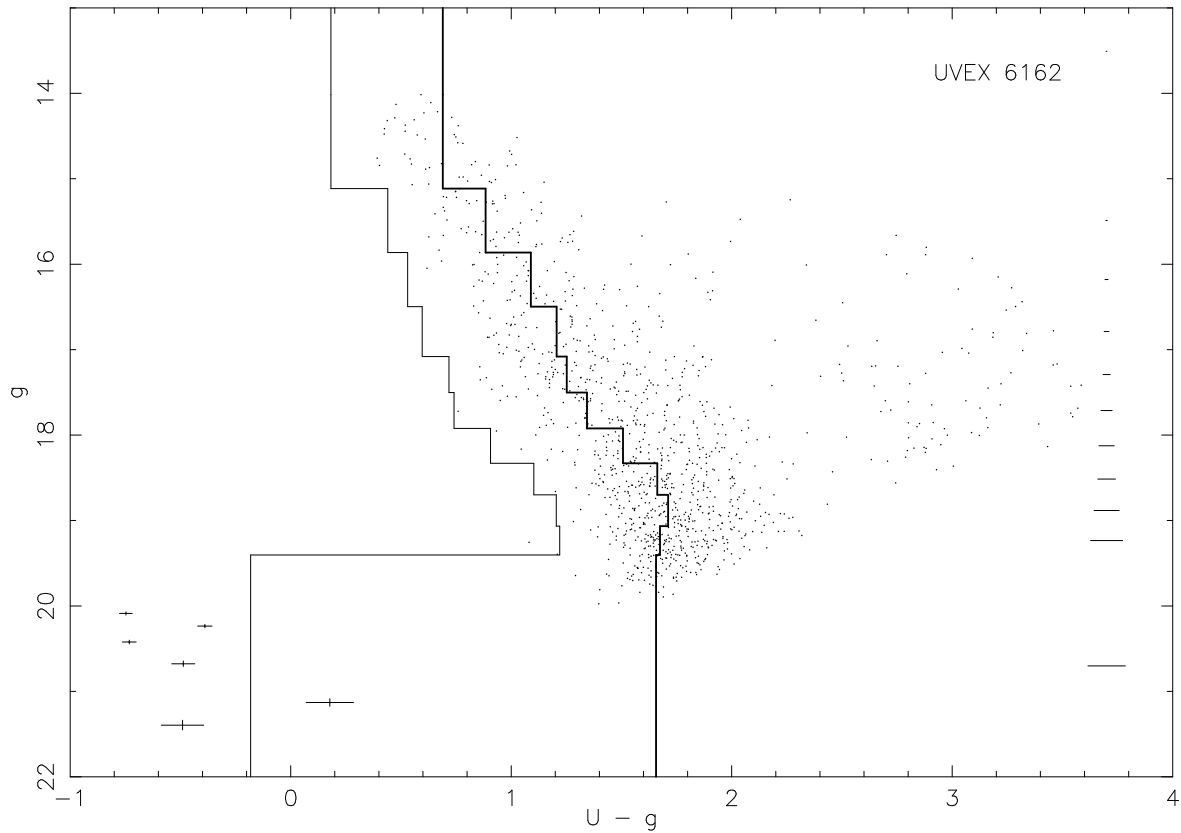}}
\caption{$g$ vs. $(U-g)$ colour-magnitude diagram of field UVEX 6162. 
The solid lines show our automated selection technique and the 
outline of the $(U-g)$ colour at which the $g$-band magnitude peaks (right, thick 
line), and the 3$\sigma$ limit for each $g$-band magnitude bin (left thin line),
called the `blue edge'. All sources more than 3 times their own photometric error to 
the left of this blue edge are selected. Not all sources plotted with their own photometric 
errors here were selected in this colour-magnitude diagram: the two blue sources fainter than 
$g$=21 are are less than 3 times their own photometric error to the left of the blue edge. 
The error bars on the right side show the average photometric error in $(U-g)$ per bin.  
\label{fig:uvex6162cmd2}}
\end{figure*}

\begin{figure*}
\centerline{\psfig{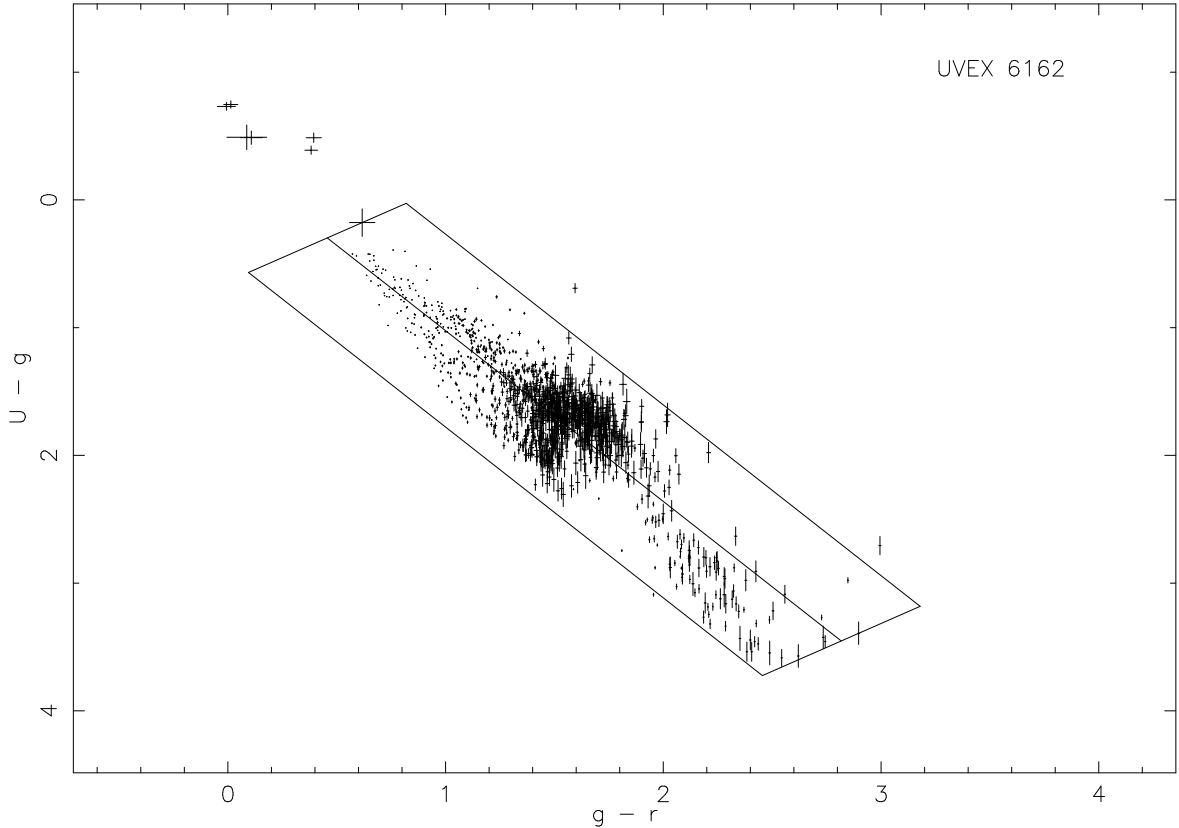}} 
\caption{$(U-g)$ vs. $(g-r)$ colour-colour diagram of field UVEX 6162. 
Data shown are all sources with a stellar or probably stellar detection in the $r$- and $g$-bands 
and with a photometric error smaller than 0.1 magnitudes in these bands. 
The lines are line fitted to the main locus (central line) 
and the 3$\sigma$ box containing the bulk of the objects.  
\label{fig:uvex6162ccd}}
\end{figure*}

One of the main aims of the \UVEX survey is to obtain a substantial and homogeneous 
sample of stellar remnants and compact binaries with well-understood selection biases. 
A large and homogeneous sample is needed to answer questions in the fields of
binary stellar evolution (e.g. on the physics of the common-envelope
phase), the earliest star formation history of our Milky Way,
the influence of chemical composition on accretion disk physics,
and the gravitational radiation foreground from compact binaries
in our Galaxy for missions such as the Laser Interferometer Space Antenna (LISA, Nelemans et al., 2004).
Stellar remnant populations to be extracted from \UVEX include
single and binary white dwarfs, subdwarf B stars, Cataclysmic Variables,
AM Canum Venaticorum (AM CVn) stars 
and neutron star and black hole binaries.
Our understanding of these populations is often limited due to the small number of 
known systems and/or strong selection biases in known samples (see e.g. Pretorius, 
Knigge \& Kolb, 2007).\\

Traditionally blue/UV surveys, targeting quasars and AGN, specifically concentrate on 
higher Galactic latitudes to avoid the problems of crowding and dust extinction.
The post-common-envelope objects mentioned above are generally  hot but intrinsically faint 
so their absolute visual magnitudes are several magnitudes lower than
main-sequence stars with similar colours. 
In the plane of the Milky Way an intrinsically blue, low-luminosity population of objects is 
visible in the $g$ vs. $(U-g)$ and $g$ vs. $(g-r)$ colour-magnitude diagrams as 
UV-excess sources against a background of higher luminosity, more distant 
and therefore more highly reddened, objects. Examples of colour-magnitude 
diagrams and colour-colour diagrams showing this effect are shown in G09 (Figs. 13 and 14) 
and Figs.\ \ref{fig:uvex6162cmd} and \ref{fig:uvex6162cmd2} above.\\

It is our aim to distinguish UV-excess sources from the normal main-sequence stars in a 
user-independent way and we have therefore developed an automated selection 
technique to construct our catalogue, presented in Sect.\ \ref{sec:algorithm}. 
In Sect.\ \ref{sec:selection} we apply this selection technique to the \UVEX data of 211 square degrees and analyse the 
spatial, magnitude and colour distributions of the UV-excess sources selected from this area. 
In Sect.\ \ref{sec:classification} we classify the UV-excess sources making use of the \IPHAS database, 
cross-matches to Simbad and synthetic colours of stellar populations as presented in G09 and Drew et al. (2005).   
Finally in Sect.\ \ref{sec:conclusions} we summarize the results and conclusions of 
our selection algorithm and catalogue of UV-excess candidates.\\

\section{Selection of UV-excess sources}
\label{sec:algorithm}
The selection algorithm is based on a field-to-field comparison of the colour of
UV-excess objects to the bulk of the main-sequence stars. Due to
variable reddening the actual selection limits in $(U-g)$ and $(g-r)$ colours 
will not be constant with Galactic position, see Sect.\ \ref{sec:conclusions}. 
The amount of reddening due to dust extinction can vary strongly 
with Galactic latitude and longitude (see e.g. Schlegel et al., 1998). 
The selection of UV-excess sources is a multi-stage process which 
will be described here.

\newpage

\begin{itemize}
\item For each \UVEX pointing we construct a $g$ vs. $(g-r)$ and $g$ vs. $(U-g)$
  colour-magnitude diagram (Figs.\ \ref{fig:uvex6162cmd} and \ref{fig:uvex6162cmd2}) 
  and a $(U-g)$ vs. $(g-r)$ colour-colour diagram (Fig.\ \ref{fig:uvex6162ccd})
  using all stellar and probably stellar objects with a photometric error $<$0.1 magnitudes 
  (see Gonz{\'a}lez-Solares et al., 2008 for the classification of sources within the \IPHAS and \UVEX surveys).
  Each object has to be detected twice per filter band: in a direct field and in an offset field 
  (not necessarily the offset field belonging to the direct field).\\

\item In the $g$ vs. $(g-r)$ and $g$ vs. $(U-g)$ colour-magnitude diagrams the main locus of objects, 
  containing mainly reddened main-sequence stars and giants, and the selection boundary 
  at the blue side of the main locus are determined in the following steps:\\ 
  $a)$ All objects detected in the $r$- and $g$-band images are divided in $g$-band magnitude bins.
  The width of the $g$-magnitude bins and the number of bins depend on the surface density of objects in the field. 
  The total number of bins for a field is $N_{bins} = \frac{1}{3}\sqrt{N_{s}}$, with $N_{s}$ the total 
  number of selected objects in the field.
  The width ($\Delta g$) of a given bin is determined by sorting of the objects
  in such a way that for all bins the product ($\Delta g \times n_{s}$) is approximately constant, 
  with $n_{s}$ the number of objects in the given bin. 
  The differences in this product between the different bins is minimized in
  an iterative process where two sources are added from neighbouring bins to the bin in 
  which the product ($\Delta g \times n_{s}$) is the smallest, and one source is taken 
  from the bin where this product is the largest
  by shifting the boundary between two bins.
  This is repeated until the differences in
  the product ($\Delta g \times n_{s}$) is minimized.
  This procedure is a compromise between using small bins
  that allow for a precise following of the main locus and using many objects
  per bin which suppresses statistical noise.\\  
  $b)$ The colour (in $(g-r)$ or $(U-g)$) at which the distribution of objects peaks for each $g$-magnitude bin is calculated. 
  This peak is found in an iterative process. 
  Per $g$-magnitude bin a sliding window over the colour distribution is used to determine the smallest window in colour that
  contains 50$\%$ of the sources. The sample of sources within this window is again searched for the smallest
  sub-window that contain 50$\%$ of the sources (i.e. 25$\%$ of the original total $n_{s}$).   
  This process is repeated until only two sources are left in the final sub-window.
  The average colour value of the last two sources is the peak 
  of the distribution in the particular $g$-magnitude bin.
  The colours where the bins peak are then smoothed by averaging over the bin itself and two adjacent magnitude bins, 
  where the bin in the middle is given a double weight. 
  This procedure gives the thick, `ragged' lines in Figs.\ \ref{fig:uvex6162cmd} and \ref{fig:uvex6162cmd2}.\\  
  $c)$ In each $g$-magnitude bin the root-mean-square deviation for the objects that lie
  to the blue of the peak is calculated using a double pass 3-sigma-clipping.
  For each bin a Gaussian distribution in $(g-r)$ or $(U-g)$ is assumed centred on the peak value found in step $(b)$. 
  The red side of the distribution is not
  taken into account since the distribution here is intrinsically wider and does not show a 
  clear cut off, as can be seen in Fig.\ \ref{fig:uvex6162cmd}. 
  The blue edge of the main locus is now placed 
  at three times the root-mean-square deviation to the blue of the peak value.
  This determines the final selection boundary or `blue edge'. 
  This is the left, `ragged' line in Figs.\ \ref{fig:uvex6162cmd} and \ref{fig:uvex6162cmd2}.\\
 
\item In the $g$ vs. $(U-g)$ and/or $g$ vs. $(g-r)$ colour-magnitude diagrams,
  sources located more than three times their own photometric error to the 
  left of the blue edge are selected.\\
  
\begin{figure}
\centerline{\psfig{figure=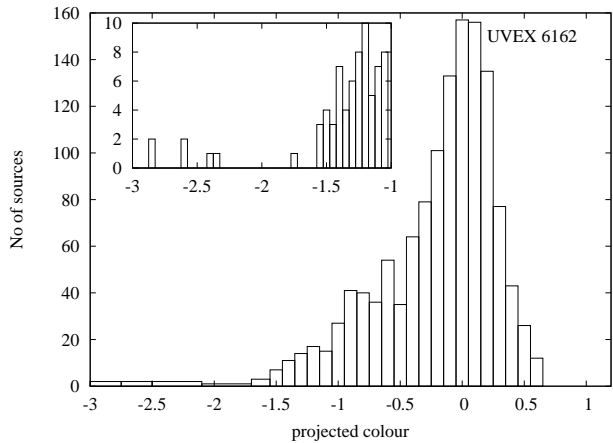,width=8.6cm,angle=0}}
\caption[]{Histogram of projected colours parallel to the 
fit line of Fig.3 for field UVEX 6162.
The upper left side of the rectangular selection box in the 
colour-colour diagram of Fig.3 is defined as the last gap in 
the first 10$\%$ of the data projected along the main locus. 
Six UV-excess sources are on the blue side of this last gap 
at `projected colour' $-1.75$ to $-2.35$.
\label{fig:ProjectedColours}}
\end{figure}

\begin{figure}
\centerline{\psfig{figure=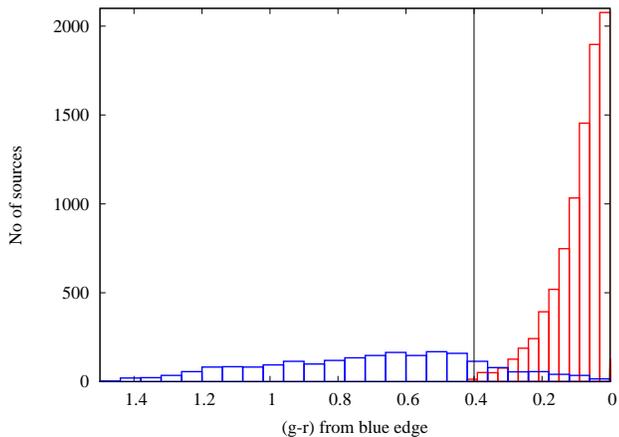,width=8.6cm,angle=0}}
\caption[]{The distribution of distances in $(g-r)$ magnitudes from the blue edge 
for objects selected from the 211 square degrees shows two populations: 
a UV-excess sample (blue) and a `subdwarf' sample (red) separated at a distance of 0.4 magnitudes from the blue edge.
The tails of these two populations partly overlap since the UV-excess 
sources at a distance less than 0.4 $(g-r)$ magnitudes from the blue edge are
at a distance more than 0.4 $(U-g)$ magnitudes from the blue edge in the $g$ vs. $(U-g)$ colour-magnitude diagram.
\label{fig:blueEdge}}
\end{figure}    

\item An additional selection is made in the $(U-g)$ vs. $(g-r)$ colour-colour diagram. 
  The average position of the bulk of all objects is 
  found by least-square fitting a straight line to 
  the distribution of all selected sources.  
  This line is broadened to a rectangular box containing the bulk of the objects (see Fig.\ \ref{fig:uvex6162ccd}).
  The fitting of this line and the broadening of this line to a selection 
  box is done in a number of steps:\\  
  $a)$ The average $(g-r)$ and $(U-g)$ colours of all selected objects are calculated. 
  The 25$\%$ of sources that lie furthest away in both colours from this average $(g-r)$,$(U-g)$ 
  position is temporarily discarded from the colour-colour diagram.\\
  $b)$ A straight line of form $(U-g)$=$m$$(g-r)$+$c$, where $m$ and $c$ are constants to be determined, 
  is fitted via least-squares to the remaining 75$\%$ of data points.
  Photometric error bars on the data points are not considered at this stage 
  to prevent the bright stars on the blue side of the main-sequence dominating the entire fit.\\ 
  $c)$ The standard deviation ($\sigma$), perpendicular to the straight line fitted to the main stellar locus, is calculated. 
  Each data points has a shortest distance $\sqrt{((\Delta(g-r))^{2}+(\Delta(U-g))^{2})}$ to the fit line, where 
  $\Delta(g-r)$ and $\Delta(U-g)$ are the differences in $(g-r)$ and $(U-g)$ colours between the fit line and the data point.
  We calculate $\sigma$ as the average of the squared differences between the fit line and all data points.
  Here the 25$\%$ discarded in the previous step is included again.
  This standard deviation is used for a sigma cut at $2\sigma$ (a cut at $3\sigma$ would discard too many main-sequence sources).
  After this sigma cut most main-sequence stars and UV-excess sources are still included in
  the sample, but most objects both blue in $(U-g)$ and red in $(g-r)$ are excluded.\\
  $d)$ A new straight line is fitted using the standard least-square method to the remaining data after the $2\sigma$ cut of step $(c)$.
  This gives the central straight line in the $(U-g)$ vs. $(g-r)$ colour-colour diagram of Fig.\ \ref{fig:uvex6162ccd}. 
  For the spread around this new line a new standard deviation ($\sigma_{final}$) is calculated. 
  $\sigma_{final}$ is the average of the squared distances between the data points and the new fit line.\\ 
  $e)$ The new standard deviation ($\sigma_{final}$) is used to broaden the fitted line to a $3\sigma_{final}$ box. 
  This gives the upper right and lower left long sides of the rectangular box in Fig.\ \ref{fig:uvex6162ccd}.\\  
  $f)$ Each data points has a projection on the fit straight line, running from the zeropoint of step $(a)$.
  These projected colours parallel to the fit line of Fig.\ \ref{fig:uvex6162ccd} are 
  shown in the histogram of Fig.\ \ref{fig:ProjectedColours}. They are used to 
  determine the location of the short upper and lower sides of the rectangular box in Fig.\ \ref{fig:uvex6162ccd}.
  Because of the typically high reddening of early-type main-sequence stars there
  usually exists a clear gap between the UV-excess sources and the main-sequence in the
  $(U-g)$ vs. $(g-r)$ colour-colour diagram, see Fig.\ \ref{fig:uvex6162ccd}. 
  This division is detected as the last gap in the first 10$\%$ of
  the data (see inset graph of Fig.\ \ref{fig:ProjectedColours}) that is larger than a threshold value
  $T = \bar d 2 ln(N)$, with $N$ the total number of data points in the field and
  $\bar d$ the average interval in projected colours between the data points.
  This works since $\bar d \propto \frac{1}{N}$ and the threshold value $T \propto \frac{ln(N)}{N}$, 
  so $T$ becomes smaller for fields with more stars.
  The factor 2 was derived by optimizing the detection of the last gap and is a compromise between
  selecting as many UV-excess sources as possible and selecting no main-sequence stars.
  In finding this gap all data points more than $3\sigma$ away from the fit line of step $(d)$ were ignored.
  For field \UVEX 6162 the last gap without sources larger than $T$ is shown in Fig.\ \ref{fig:ProjectedColours} and 
  is located at --1.75 to --2.35.
  In field \UVEX 6162 there are 6 UV-excess sources to the blue of the last gap.  
  Dense fields usually have less reddening than sparse
  fields and there are simply more stars, causing both the intervals between
  successive main-sequence stars and the gap between the main-sequence and the
  UV-excess sources to shrink. Because the number of intervals between main-sequence
  stars increases for dense fields the chance of finding a very long interval
  caused by statistical noise decreases and therefore the logarithmic term
  was added. 
  The resulting selection box around the main-sequence, starting at the 
  end of the gap and with a half-width of $3\sigma$, can be seen in Fig.\ \ref{fig:uvex6162ccd}.
  The lower boundary of the box is the reddest main locus object in the field.
  The location of the mean stellar locus and the resulting selection box are completely determined by
  the data. No theoretical requirements are desired due to the absence of a global calibration at this moment, 
  analogous to the selection method in Withham et al. (2008).\\ 
  $g)$ Fig.\ \ref{fig:uvex6162ccd} shows the fit to the main locus and the $3\sigma$ wide broadened 
  rectangular box containing the bulk of the objects of field \UVEX 6162. 
  All sources more than 3 times their own photometric error to the blue outside 
  the selection box are selected. Additionally, sources both blue in $(U-g)$ and red in 
  $(g-r)$ that are on the right side above the rectangular selection box are 
  labelled as `purple' candidates, see Sect.\ \ref{sec:selection}.\newline
  
\item In the $g$ vs. $(g-r)$ colour-magnitude diagram we put in the additional limit 
  that any UV-excess source should be bluer than $(g-r)$=1.25, 
  since all fields will contain a number of unreddened 
  M-dwarfs and the latest M-dwarfs have $(g-r)$$\sim$1.25 (see G09).
  We found that for some highly obscured fields these unreddened M-dwarfs are actually the bluest 
  stars in the field of view, see for example field UVEX 6167 in Fig.14 of G09. 
  Any object that meets the criteria listed above is selected.\\
  
\item Selected blue objects are further separated on the basis of their distance 
  in colour from the blue edge in the colour-magnitude diagrams. 
  Fig.\ \ref{fig:blueEdge} shows the distribution of distances in $(g-r)$ magnitudes from the blue edge 
  of the sources selected on the left side of the blue edge.
  These sources were selected from the first 211 square degrees of \UVEX data, see Sect.\ \ref{sec:selection}.
  The distribution in Fig.\ \ref{fig:blueEdge} shows two populations that overlap at 0.4 magnitudes away from the blue edge. 
  Despite the overlap we label the objects that are more than 0.4 magnitudes away from the blue edge of the main-sequence 
  in $g$ vs. $(U-g)$,  $g$ vs. $(g-r)$ or in both colour-magnitude diagrams as `UV-excess' candidates, and 
  we label sources less than 0.4 magnitudes away from the blue edge as `subdwarf' candidates.
  In Fig.\ \ref{fig:blueEdge} the two populations do overlap, 
  because the UV-excess candidates closer than 0.4 magnitudes from the blue edge 
  are selected in the $g$ vs. $(U-g)$ colour-magnitude diagram.
  Objects that were selected in the blue corner of the $(U-g)$ vs. $(g-r)$ colour-colour diagram
  are labelled as `UV-excess' candidates when they are more than 0.4 magnitudes away from the blue edge
  in both or one of the colour-magnitude diagrams.
  Objects that were only selected in the blue corner of the $(U-g)$ vs. $(g-r)$ colour-colour diagram
  and less than 0.4 magnitudes away from the blue edge in the colour-magnitude diagrams are labelled as `subdwarf' candidates.
  This is to avoid that scattered main-sequence sources that are just outside the selection box 
  are added to the UV-excess catalogue (see Sect.\ \ref{sec:conclusions}).
  The subdwarf population, closest to the main-sequence, can be reddened main-sequence stars if the 
  spatial scale of the reddening strongly varies within the field of view of one pointing.
  Another likely explanation is that these objects are a
  population of unreddened metal-poor stars, which can have absolute magnitudes
  fainter by up to two magnitudes compared with main-sequence stars of
  the same colour (e.g. Johnson, 1955). Their smaller distances, up to a factor 2.5 closer
  compared to normal main-sequence stars of the same colour, 
  will subject them to a smaller reddening. They will therefore stand out
  in the $(g)$ vs. $(g-r)$ colour-magnitude diagram.
  We therefore tentatively label this population as `subdwarfs'.\\
\end{itemize}

\begin{figure*}
\centerline{\epsfig{figure=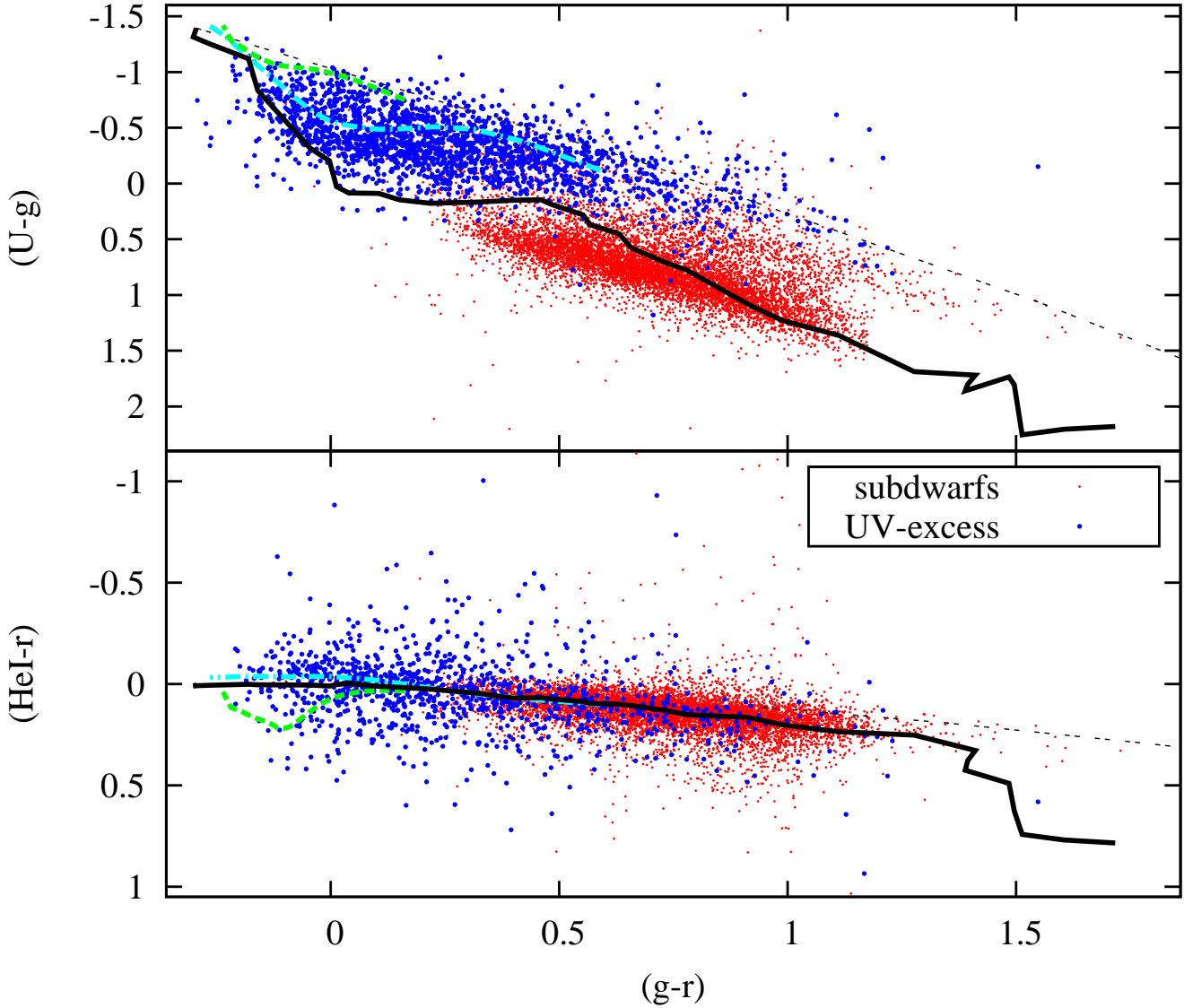,width=27cm,clip=}}
\caption[]{UVEX colour-colour diagrams with all UV-excess candidates (blue) and 'subdwarfs' (red) 
selected from the first 211 square degrees of UVEX data. Overplotted are the new simulated unreddened main-sequence (solid black) and the 
O5V- and supergiant reddening lines (dashed black). 
These new synthetic colours are different from the colours in G09 since they include the effect of the $U$-band 
filter red leak (see Appendix B). The cyan and green dashed lines are the simulated unreddened Koester DA and DB white dwarf tracks.
Since a global photometric calibration was not applied yet the colours plotted here might be scattered a bit. 
Spectroscopic follow-up of the sources around $(HeI-r) < -0.2$ in the plot above which lay more than 3$\sigma$ 
their photometric error above the synthetic DA white dwarf track shows that these objects are 
mainly scattered white dwarfs (Verbeek et al., 2011b, in prep). 
\label{fig:UVexcessSampleCCD}}
\end{figure*}

\begin{figure*}
\centerline{\epsfig{figure=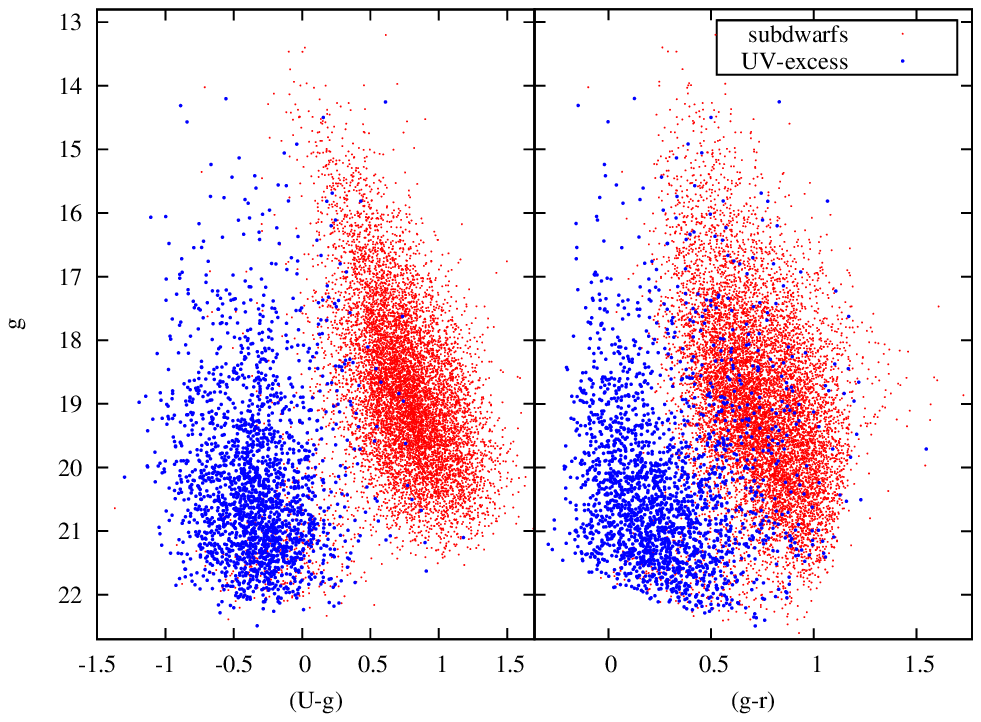,width=22cm, clip=}}
\caption[]{UVEX colour-magnitude diagrams with all UV-excess candidates (blue) and 'subdwarfs' (red) 
selected from the first 211 square degrees of UVEX data.
Since a global photometric calibration was not applied yet the colours plotted here might be scattered a bit.  
\label{fig:UVexcessSampleCMD}}
\end{figure*}

\section{Applying the selection method: a first sample of UV-excess sources}
\label{sec:selection}
The selection algorithm described in Sect.\ \ref{sec:algorithm} and displayed in 
Figs.\ \ref{fig:uvex6162cmd} to \ref{fig:blueEdge} 
is applied to all `good' \UVEX data up to November 2008. 
These data were obtained in June 2006, May 2007, July 2007, September 2007 and October 2007 
(the data of October 2007 have no $\he$ detections). 
Here `good' data are defined as a $g$-band seeing $<$1\farcs7 and 
an $r$-band background count rate smaller than 2000 ADU (a high background 
count rate indicates observations close to the Moon or clouds).
Thirty-three fields with large photometric shifts in $(g-r)$, $(U-g)$ or $(HeI-r)$ were excluded.
This results in a total of 726 direct fields (i.e. 211 square degrees).
In the $g$- and $r$-bands only `stellar' and `probably stellar' sources are allowed. 
In the $U$- and $\he$ bands we also allow sources with morphology class 0 (`noise-like'), 
since the longer integration times for $U$ and $\he$ sometimes caused trailing.\\

The number of all selected stellar sources in the $g$- and $r$-bands from these 
211 square degrees is $7 \times 10^{6}$ (on average $3.3 \times 10^{4}$ per square degree).
The number of objects that have a $U$-band detection is $1.5 \times 10^{6}$ (22$\%$), 
the number of objects with a $\he$ band detection is $3 \times 10^{6}$ (41$\%$)  
and the number of objects with both a $U$- and $\he$ band detection is $9 \times 10^{5}$ (13$\%$).
There are $6.3 \times 10^{5}$ sources that have a $U$-band detection but no $\he$ detection.
These sources are clearly visible in the $\he$ images but have morphology class 0 (`noise-like'), 
which we allowed in the $U$- and $\he$ bands. 
As a final check the \UVEX finder charts of sources that pass the selection algorithm were eye-balled. 
Blended objects, sources with noise, double sources and sources close to a bright star are removed.
The eye-balling was done in a conservative way: sources with a small artefact which still might be okay were also removed. 
Only 3$\%$ of the UV-excess candidates was removed in the eye-balling. Five percent of the sources 
was removed because of duplicate entries in the catalogue due to field overlap. 
This is consistent with the overlap between direct fields and overlap fields (Gonz{\'a}lez-Solares et al., 2008).
The total number of unique sources that pass the UV-excess selection is 
2\,170 UV-excess sources and 9\,872 additional `subdwarfs'. 
The numbers in this paragraph are summarized in Table \ \ref{tab:stellar}.\\

All UV-excess candidates and `subdwarf' sources are plotted in the colour-colour and 
colour-magnitude diagrams of Figs.\ \ref{fig:UVexcessSampleCCD}
and \ref{fig:UVexcessSampleCMD}. A global photometric calibration is not applied to the \UVEX data 
yet, so the magnitudes and colours of the UV-excess sources plotted in 
Figs.\ \ref{fig:UVexcessSampleCCD} and \ref{fig:UVexcessSampleCMD} might show a small scatter.
The synthetic colours in the colour-colour diagrams are different from the simulated
colours in G09 since they include the effect of the $U$-band 
filter red leak (see Appendix\ \ref{app:redleakuvexcolours}).
In the colour-magnitude diagrams and in the $(U-g)$ vs. $(g-r)$ colour-colour diagram
it can be seen that the UV-excess candidates and the subdwarf sample separate.\\
 
Of the objects in the UV-excess catalogue 
98$\%$ is selected in $g$ vs. $(g-r)$,
66$\%$ is selected in $g$ vs. $(U-g)$,
82$\%$ is selected in $g$ vs. $(g-r)$ more than 0.4 magnitude from the blue edge,
63$\%$ is selected in $g$ vs. $(U-g)$ more than 0.4 magnitude from the blue edge,
44$\%$ is selected in both colour-magnitude diagrams more than 0.4 magnitude from the blue edge,
and 63$\%$ is selected in the $(U-g)$ vs. $(g-r)$ colour-colour diagram.
Of the UV-excess candidates 
34$\%$ is selected 0.4 magnitude from the blue edge in $g$ vs. $(g-r)$ but completely not in $g$ vs. $(U-g)$ and 
2$\%$ is selected 0.4 magnitude from the blue edge in $g$ vs. $(U-g)$ but completely not in $g$ vs. $(g-r)$.
Without the $U$-band filter 16$\%$ of the sources would not be selected in the UV-excess catalogue.\\

The number of UV-excess objects that have a `stellar' or `probably stellar' $U$-band detection is 1546 (71.2$\%$), 
the number of objects with a `stellar' or `probably stellar' $\he$ band detection is 667 (30.7$\%$)
and the number of objects with both a `stellar' or `probably stellar' $U$-band and $\he$ band detection is 
513 (23.6$\%$).\\

Surprisingly we also find a third fairly large population of
objects that are blue in $(U-g)$ as well as red in $(g-r)$. 
All these sources are selected in the upper right half of the 
$(U-g)$ vs. $(g-r)$ colour-colour diagram (see Fig.\ \ref{fig:uvex6162ccd}), 
outside the rectangular box fitted to the main-sequence and are labelled as `purple' 
(i.e. red and blue) sources.
Purple sources in the upper right half of the $(U-g)$ vs. $(g-r)$ colour-colour diagram that are below the 
O5V-reddening line (see Fig.\ \ref{fig:UVexcessSampleCCD} and Fig.6 of G09) are deselected 
because they can be reddened main-sequence stars.
Of the sources in the purple sample 100$\%$ is selected in the $(U-g)$ vs. $(g-r)$ colour-colour diagram, 
6$\%$ is also selected in $g$ vs. $(U-g)$ more than 0.4 magnitude from the blue edge and
0$\%$ is also selected in $g$ vs. $(g-r)$ more than 0.4 magnitude from the blue edge.
These sources more than 0.4 magnitude from the blue edge are not in the UV-excess catalogue.
Of the purple sources 87$\%$ is selected only in the $(U-g)$ vs. $(g-r)$ colour-colour 
diagram but not at all in a colour-magnitude diagram.\\

\begin{table}
\caption[]{All (probably) stellar sources from 211 sq.deg. \label{tab:stellar} }
{\small
\begin{tabular}{ | l | r | c | }
    \hline
Objects detected:                     &     Number:    	     &  Fraction ($\%$):   \\ \hline
in the $g$ and $r$-band               &  $7 \times 10^{6}$   &    100 \\
in the $U$-band                       &  $1.5 \times 10^{6}$  &     22 \\
in the $HeI$-band                     &  $3 \times 10^{6}$   &      41 \\
in the $U$ and $HeI$-band             &  $9 \times 10^{5}$  &     13 \\
after UV-excess selection             &      2\,170         &    0.03 \\	      
    \hline
\end{tabular} \\ 
}
\end{table}

\begin{figure}
\centerline{\epsfig{figure=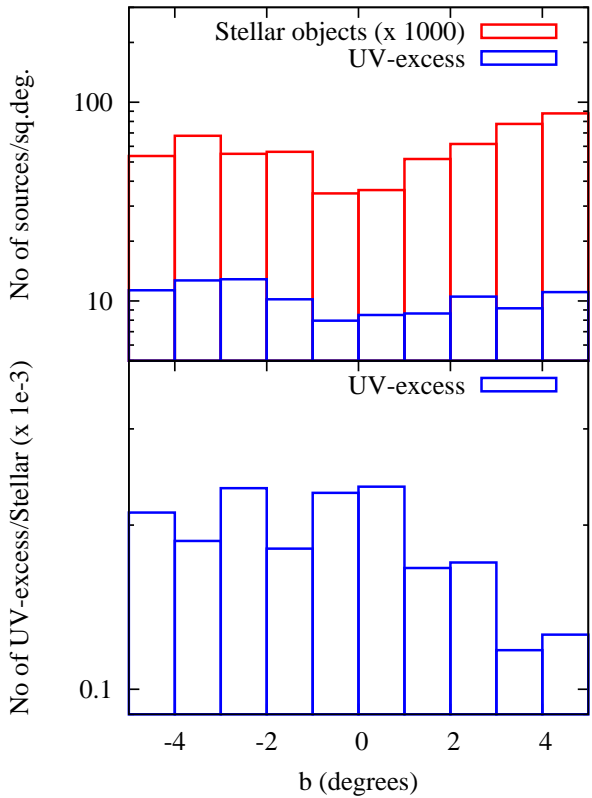,width=16cm}}
\caption[]{Distribution of stellar sources and UV-excess sources over Galactic latitude (upper graph). 
The y-axis (in logscale) shows the number of stellar sources and UV-excess candidates per square degree for the associated Galactic latitude bin.
The bottom graph shows the ratio of UV-excess candidates and stellar sources distributed over Galactic latitude. 
\label{fig:galacticdistribution}}
\end{figure}

\subsection{The properties of the selected UV-excess catalogue}
The complete UV-excess catalogue and the two additional `subdwarf' and `purple' samples can be obtained 
from the \UVEX website\footnote{http://www.uvexsurvey.org} 
or via VizieR\footnote{http://vizier.u-strasbg.fr/viz-bin/VizieR}. 
The columns of the catalogue and the different selection labels are explained in Appendix\ \ref{app:uvexcesscatalogue}.\\

Both the stellar sources and UV-excess candidates selected from the 211 square degrees of \UVEX data 
are non-uniformly distributed over Galactic latitude and longitude. 
Fig.\ \ref{fig:galacticdistribution} shows the distributions of all stellar objects and UV-excess 
objects over Galactic latitude. The surface density of all stellar  
sources in the $r$- and $g$-bands shows a minimum centred on the Galactic equator. 
The distribution of the UV-excess sources is not directly straightforward and depends on Galactic latitude.

\section{Source classification and cross-matches with other catalogues}
\label{sec:classification}
In Sect.\ \ref{sec:selection} we have selected the UV-excess catalogue 
and two additional samples: a `subdwarf' sample and a  `purple' sample. 
Despite these names each class is expected to consist of a mix of populations. 
For instance the UV-excess sample will contain white dwarfs,
Cataclysmic Variables, AM CVn stars and central stars of planetary nebulae. 
We have therefore cross-correlated our samples with known source catalogues, 
the results of which are shown in Figs.\ \ref{fig:SimbadXmatch} 
to\ \ref{fig:iphasXmatch} and summarized in Table \ \ref{tab:crossmatching}.
Spectroscopic follow-up has also been obtained and will be presented 
in Verbeek et al., 2011b (in prep.).\\  

\subsection{Matches with the Simbad database}
\begin{figure}
\centerline{\epsfig{figure=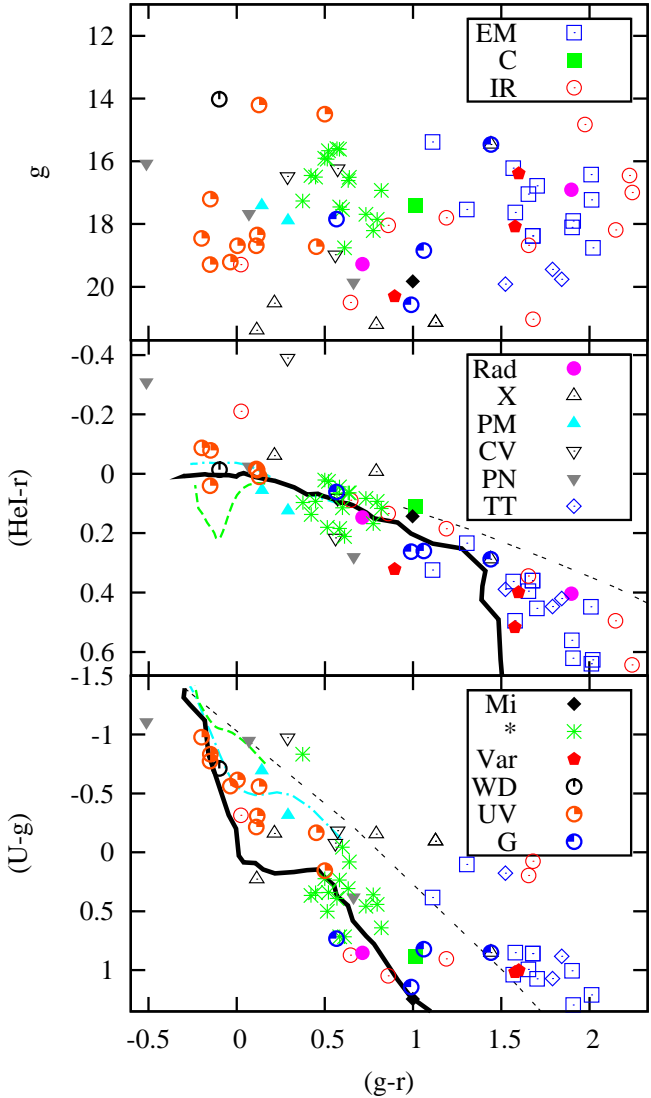,width=22cm}} 
\caption[]{Matches between the UV-excess catalogue and Simbad plotted in the \UVEX
colour-colour and colour-magnitude diagrams. The black solid lines are the simulated colours of unreddened main-sequence stars, 
the dashed lines are the unreddened DA and DB white dwarf tracks (cyan and green) and the O5V- and supergiant reddening line (black dashed).
A part of the matches between the additional subdwarf and purple samples and Simbad are also plotted.
The legend continues in the three sub-panels. The Simbad object types plotted here are: Emission-line star (EM), Carbon star (C),
Infra-Red source (IR), Radio-source (Rad), X-ray source (X), High proper-motion star 
(PM), Cataclysmic Variable star (CV), Planetary Nebula (PN), T Tau-type star (TT),
Mira-type star (Mi), star (*), Variable star (Var), White dwarf (WD), UV-emission source (UV) and Galaxy (G). 
\label{fig:SimbadXmatch}}
\end{figure}

There are 24 matches (only $\sim1\%$ of all UV-excess sources) between the 
UV-excess catalogue and known sources in Simbad, using a matching radius
of 10 arcsec. These matches are plotted with their 
\UVEX colours and Simbad classifications in the  
$(U-g)$ vs. $(g-r)$ and $(HeI-r)$ vs. $(g-r)$ colour-colour and $g$ vs. $(g-r)$ colour-magnitude diagrams 
of Fig.\ \ref{fig:SimbadXmatch}. 
The UV-excess candidates that are found in Simbad include 2 planetary nebulae, 3 Cataclysmic Variables, 1 DA white dwarf, 
2 high proper motion sources, 2 X-ray sources, 1 emission line star, 10 UV sources, 1 IR source and 2 seemingly normal stars.\\ 

Additionally there are 26 known objects from the subdwarf sample in Simbad: 1 planetary nebula, 1 X-ray source, 1 UV source, 1 Mira star, 1 Carbon star, 
2 IR sources, 1 radio source, 3 galaxies, 2 variable stars and 13 seemingly normal stars. There are 87 objects from the purple sample in Simbad: 
2 X-ray sources, 4 T Tauri stars, 6 Mira stars, 5 Carbon stars, 12 emission line stars, 44 IR sources, 1 radio source, 2 galaxies, 3 variable stars 
and 8 seemingly normal stars. 
The fact that three planetary nebulae are found in Simbad shows that in the \UVEX data the central star is visible. 

\begin{figure}
\centerline{\epsfig{figure=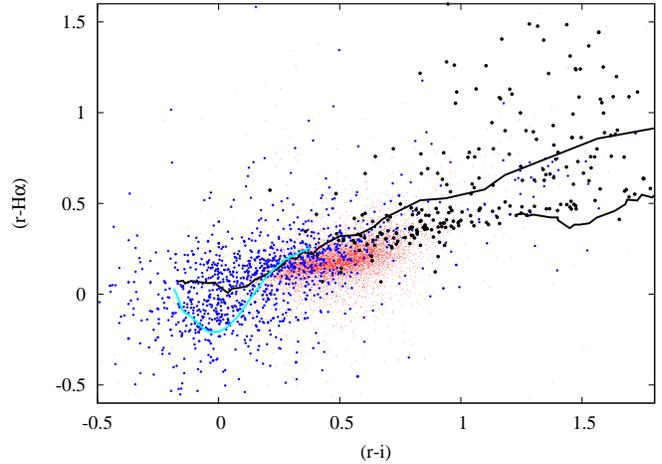,width=9.2cm,angle=0}}
\caption[]{All matches between the UV-excess catalogue (blue dots), the two additional subdwarf (red) and purple (black) samples 
and the IPHAS Initial Data Release plotted in the IPHAS (r-H$\alpha$) vs. $(r-i)$ colour-colour diagram.  
Overplotted are the simulated main-sequences with $E(B-V)=0$ and $E(B-V)=2$ (black solid lines) 
and the unreddened Koester DA white dwarf track (cyan).
\label{fig:iphasIDR}}
\end{figure}

\subsection{$\IPHAS$: The Deacon, Corradi, Viironen and Witham catalogues}
The INT/WFC Photometric H$\alpha$ Survey of the Northern Galactic Plane (\IPHAS)
(Drew et al., 2005) images the same survey area as \UVEX with the same telescope and camera set-up
using the $r$, $i$ and $H\alpha$ filters. 
The astrometric precision of \IPHAS is better than 0.1 arcsec (Gonz{\'a}lez-Solares et al., 2008) 
and on the same frame as \UVEX. 
The \IPHAS initial data release (IDR) covers approximately 89 percent of the total survey area. 
The result of the cross-matching with a match radius of 1.0 arcsec between our UV-excess catalogue and the \IPHAS 
IDR is displayed in the $(r-H\alpha)$ vs. $(r-i)$ colour-colour diagram of Fig.\ \ref{fig:iphasIDR}.
Although the match radius of 1.0 arsec is much larger than the astrometric precision of both surveys, it
makes sure that all matches are found while it is still too small for mismatches 
(see Gonz{\'a}lez-Solares et al., 2008, Fig.\,6). Close neighbours and ambiguities in the UV-excess 
catalogue were already removed in the eye-balling process.\\

There is a match in the \IPHAS IDR for 1203 objects of the UV-excess catalogue. 
The majority of the UV-excess candidates with an \IPHAS IDR match overlap with the 
position of the simulated unreddened DA white dwarf 
tracks in the $(r-H\alpha)$ vs. $(r-i)$ colour-colour diagram
of Fig.\ \ref{fig:iphasIDR}. Here a part of the UV-excess candidates
is reddened and scattered. These objects might also be QSOs and Cataclysmic Variables.\\ 

Additionally 8177 objects of the subdwarf sample and 
565 objects of the purple sample have a match in the \IPHAS IDR. 
The subdwarfs are at the position of slightly reddened ($E(B-V)<1$) A-type stars
in the $(r-H\alpha)$ vs. $(r-i)$ colour-colour diagram of Fig.\ \ref{fig:iphasIDR}.
A small fraction of the purples is saturated in the $i$-band 
\IPHAS IDR data ($\sim20\%$ of the purples that do not have an \IPHAS IDR match), 
an other fraction is blended in the \IPHAS images.
In the $(r-H\alpha)$ vs. $(r-i)$ colour-colour diagram it becomes clear that 
the purple sample separates out in several sub-samples: 
one or more is located below the position of the K- and M-dwarfs of the simulated 
unreddened main-sequence colours at $(r-H\alpha)\sim 0.4$ and $(r-i)\sim 1$, 
a second population is located below the position of the late type stars 
at reddening $E(B-V)\sim 2$ around $(r-H\alpha)\sim 1$ and $(r-i)\sim 3$. 
There is possibly a third population of H$\alpha$ emission line objects.\\

From \IPHAS several catalogues were selected for high-proper motion objects, H$\alpha$ emission line stars, 
Symbiotic stars and planetary nebulae candidates. The result of the cross-match between these catalogues
and the UV-excess catalogue is shown in the next subsections. 

\begin{figure}
\centerline{\epsfig{figure=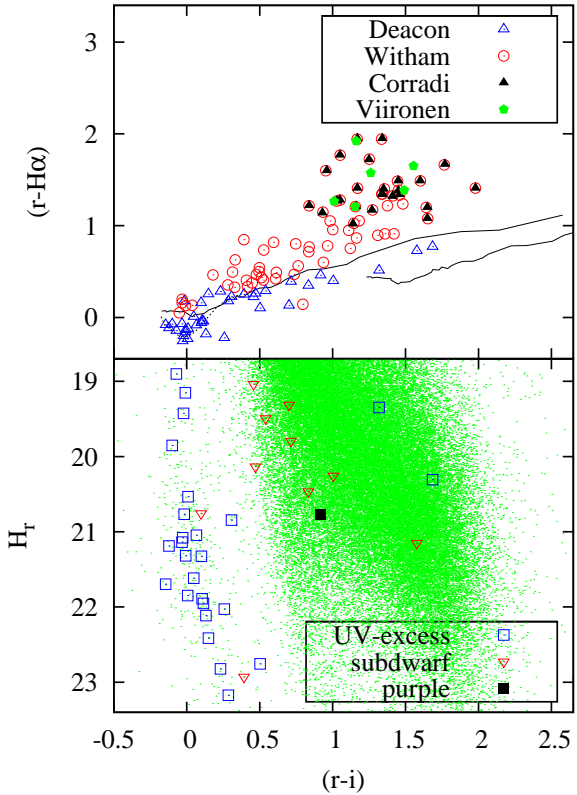,width=16cm}}
\caption{IPHAS $(r-H\alpha)$ vs. $(r-i)$ colour-colour diagram (upper graph) with the matches between 
the UV-excess catalogue, the subdwarf and purple samples 
and the IPHAS-POSS1 catalogue (blue open triangles), Witham H$\alpha$ emitters (red circles), 
Corradi symbiotic candidates (black triangle) and Viironen planetary nebulae (green diamonds).
The black solid lines are the main-sequence at E(B-V)=0 and E(B-V)=2, the dashed 
line is the unreddened synthetic Koester DA white dwarf track.
The bottom graph shows all objects of the IPHAS-POSS1 proper motion catalogue (green dots) 
in the reduced proper motion diagram and the matches between the UV-excess catalogue and the IPHAS-POSS1 catalogue 
overplotted. The blue, red and black points are respectively matches with the UV-excess catalogue, subdwarf sample and purple sample.  
There is one additional purple source at $(r-i)$=3.4, $(r-H\alpha)$=0.9, $H_{r}$=19.
\label{fig:DeaconPM}}
\end{figure}

\subsubsection{High-proper motion systems: IPHAS-POSS1 proper motion stars}
The IPHAS-POSS1 Proper Motion catalogue (Deacon et al., 2009) is constructed from 
POSS-I Schmidt plate data taken in the 1950s and 1400 square degrees of \IPHAS Galactic Plane Survey data 
as a second epoch. This proper motion catalogue has roughly a fifty years baseline and contains
populations such as main-sequence stars, white dwarfs, subdwarfs and Cataclysmic Variables.
The catalogue contains $\sim10^{5}$ objects with proper motions smaller than
150 millarcseconds per year in the magnitude range $13.5< r <19$.
In the IPHAS-POSS1 Proper Motion catalogue there is a match for 26 UV-excess candidates.
We used a matching radius of 1.0 arcsec because of the \IPHAS-\UVEX $r$-band baseline 
of a few years, but all matches were within a radius of 0.4 arcsec. 
The UV-excess candidates that are in the IPHAS-POSS1 
catalogue have proper motions between 15 mas/yr and 150 mas/yr. 
The matches are overplotted in the reduced proper motion ($H_{r}$) diagram 
and \IPHAS colour-colour diagram of Fig.\ \ref{fig:DeaconPM} together
with all the objects in the Deacon IPHAS-POSS1 catalogue. 
Here reduced proper motion is defined as $H_{r} = r + 5 log(\mu) + 5$, 
where $r$ is the $r$-band magnitude and $\mu$ is the proper motion 
in arcseconds per year (see Deacon et al., 2009).
The majority of the proper motion UV-excess sources overlaps the 
white dwarf population around $(r-i)\sim0$.
In the $(r-H\alpha)$ vs. $(r-i)$ colour-colour diagram the UV-excess sources mainly overlap 
with the simulated unreddened white dwarfs track. Also in the \UVEX colour-colour diagrams of Fig.\ \ref{fig:iphasXmatch} 
the matches with the IPHAS-POSS1 catalogue overlap with simulated white dwarfs tracks.
Additionally there is a match for 9 objects from the subdwarf sample and 2 objects from the purple sample.

\subsubsection{H$\alpha$-emission line systems: Witham IPHAS H$\alpha$ emission-line objects}
The Witham catalogue of H$\alpha$ emission-line objects (Witham et al., 2008) is extracted 
from $\sim$80 percent of the \IPHAS Galactic Plane Survey area data.
This catalogue, selected from the $(r-H\alpha)$ vs. $(r-i)$ colour-colour diagram, contains 4853 point sources clearly showing H$\alpha$-excess.
Spectroscopic follow-up of $\sim$300 candidates shows that more than 95 percent
are indeed true H$\alpha$ emitters such as early-type emission-line stars, active late-type stars, interacting binaries,
young stellar objects and compact nebulae (see Wesson et al., 2008, Viironen et al., 2009).
When the UV-excess catalogue is cross-matched with the Witham catalogue 
there is a match within a radius of 1.0 arcsec for 15 UV-excess candidates and additionally there is a match for 
15 objects in the subdwarf sample and 37 objects in the purple sample. These matches are plotted in 
Figs.\ \ref{fig:DeaconPM} and \ \ref{fig:iphasXmatch}.

\subsubsection{Symbiotic stars: Corradi IPHAS symbiotic catalogue}
Symbiotic stars (see e.g. Corradi et al., 2008 and 2010) are interacting binaries composed of a hot white dwarf, 
accreting from a cool giant companion through a wind or through Roche lobe overflow. 
Two types ``Stellar'' (S) and ``Dusty'' (D) exist with orbital periods 
between 1-16 years, and more than 20 years, respectively. 
The number of known systems in our Galaxy, where 80 percent is S-type, is roughly 200 including 26 suspected candidates,
but the size of the total symbiotic population in the Milky Way is still poorly known (Belczy\'nski et al., 2000).
The \IPHAS symbiotic catalogue (Corradi et al., 2008) was extracted from \IPHAS Galactic Plane Survey data. 
Cross correlating the \IPHAS symbiotic catalogue (Corradi et al., 2008) with the UV-excess 
catalogue results in a match for 2 sources within a radius of 1.0 arcsec.
Additionally there is a match for 23 purple sources. These matches are also in the Witham H$\alpha$ emission line star catalogue. 
This large fraction of purple matches is due to the position 
of the symbiotic stars in the \UVEX colour-colour diagrams, see Fig.\ \ref{fig:iphasXmatch}.
In the $(HeI-r)$ vs. $(g-r)$ colour-colour diagram the symbiotic candidate matches are to 
the right of the characteristic ``hook'' of the simulated unreddened main-sequence colours.
Note that a fairly large fraction of the sources in the \IPHAS symbiotic catalogue (Corradi et al., 2008) are likely to be young
stellar objects.
For most of the symbiotic candidate matches the available spectroscopic data were checked. 
The majority of the \UVEX symbiotic candidate matches turn out to be classical T Tauri stars (Verbeek et al., 2011b, in prep.) while 
none of them is a known symbiotic star, and they are found in regions of the sky close to other H$\alpha$ emitters.
When the \UVEX colours of these matches are compared with simulated symbiotic colours 
the colours of the symbiotic candidate matches are too blue.
Therefore, these matches are not good symbiotic star candidates
as symbiotic stars in the Galactic Plane are generally redder.

\subsubsection{Planetary Nebulae: Viironen IPHAS Planetary Nebulae}
\begin{figure}
\centerline{\psfig{figure=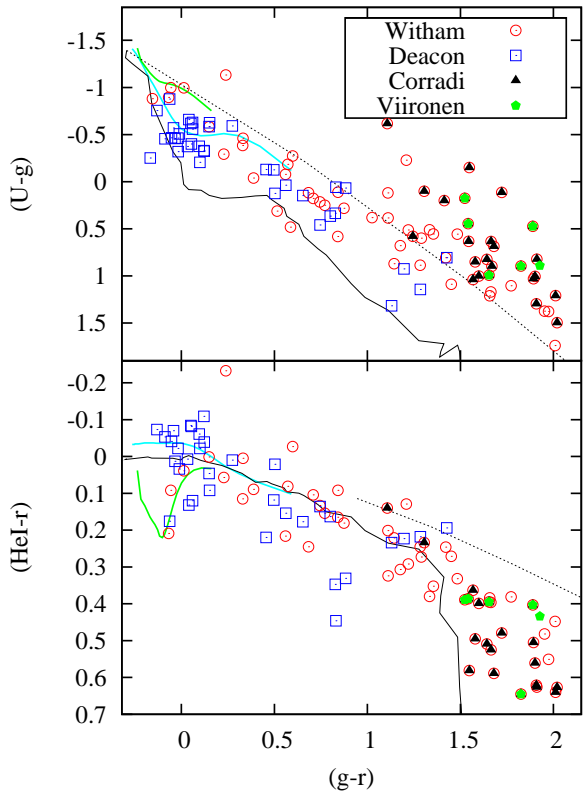,width=16cm}}
\caption{Result of the cross-matching between the UV-excess catalogue, the subdwarf and purple samples and the 
IPHAS H$\alpha$ emission line objects (Witham), IPHAS-POSS1 high proper motion stars (Deacon), IPHAS symbiotic candidates (Corradi) and 
IPHAS planetary nebula candidates (Viironen), overplotted on the simulated unreddened main-sequence (solid black line) and 
the unreddened DA and DB white dwarf tracks (cyan and green lines). The dashed lines are the O5V-reddening line (upper graph) and 
the supergiant reddening line (bottom graph).  
There is one additional purple source at $(g-r)$=3.02, $(U-g)$=2.22, $(r-HeI)$=1.19. 
\label{fig:iphasXmatch}}
\end{figure}

Approximately 2700 planetary nebulae, the evolutionary product of 0.8--8\msun 
main-sequence stars before they become a white dwarf, have been found in the Milky Way so far while 
population syntheses predicts a larger population ($4.6 \pm 1.3$ Galactic PN with radii $<0.9\,pc$, Moe et al., 2006). 
Extinction makes it difficult to observe planetary nebulae in the direction of the Galactic Plane.
The Viironen planetary nebula catalogue (Viironen et al., 2009) contains 
781 planetary nebula candidates selected from the \IPHAS data. 
The list includes very young and proto-planetary nebulae, normal 
planetary nebulae and other non-planetary nebula emission line objects.
Cross-correlating the Viironen catalogue with the UV-excess catalogue results in 0 matches within 1.0 arcsec. 
Only from the purple sample 6 objects have a match with the Viironen catalogue.
These planetary candidate matches are in the same regime as the symbiotic candidates matches. 
Their location in the \IPHAS and \UVEX colour-colour diagrams is shown 
in Figs.\ \ref{fig:DeaconPM} and \ \ref{fig:iphasXmatch}. 
None of the matches is a known planetary nebula, and due to the blue \UVEX colour of these 
matches they are not good planetary nebulae candidates.

\subsection{The Two-Micron All-Sky Survey}
The selected UV-excess sample is cross-matched with the Two-Micron All-Sky Survey (2MASS) data
(Cutri et al., 2003). Of the UV-excess candidates 171 objects can be found in 2MASS within a radius of 1.0 arcsec. 
Additionally there is a match in 2MASS for 3605 subdwarfs and 732 purple sources. 
This shows the good match of \UVEX with 2MASS, in particular for the redder sources.  
The cross-match between the UV-excess catalogue, subdwarf and puple samples 
and 2MASS will be further discussed in Verbeek et al., 2011c (in prep.).\\

\subsection{The UKIRT InfraRed Deep Sky Surveys}
Finally the UV-excess catalogue is cross-matched with the the UKIDSS (UKIRT InfraRed Deep Sky Surveys) 
Galactic Plane Survey (Lucas et al., 2008). This survey images the Northern Galactic plane in the same Galactic
latitude range as UVEX in the filters $J$, $H$ and $K$. There is a match for 576 UV-excess candidates within a radius of 1.0 arcsec 
in one of the filters. Of these UV-excess candidates 219 sources have a matches in all three filters. 
Additionally there is a match for 4208 subdwarfs and 506 purples. 
The cross-match between our samples and the UKIDSS GPS will also be further 
discussed in Verbeek et al., 2011c (in prep.).\\ 

\begin{table}
\caption[]{Summary of the cross-matching. \label{tab:crossmatching} }
{\small
\begin{tabular}{ | l | l | }
    \hline
Catalogue:                 &  Nr and fraction ($\%$) of UV/s/p:     \\ \hline
Simbad                     &  24/26/87 (1.11/0.26/10.8)           \\		 
2MASS                      &  171/3\,605/732 (7.9/37/91)          \\
UKIDSS                     &  576/4\,208/516 (26/43/64)           \\
IPHAS IDR                  &  1\,203/8\,177/565  (55/83/70)       \\
Deacon IPHAS-POSSI         &  26/10/2  (1.20/0.10/0.25)    	 \\
Witham H$\alpha$           &  15/15/37  (0.69/0.15/4.61)  	 \\
Corradi Symbiotics         &  2/0/23  (0.09/0.00/2.86)    	 \\
Viironen PNs               &  0/0/6  (0.00/0.00/0.75)	   	 \\		
    \hline
\end{tabular} \\ 
}
\end{table}

\section{Discussion and conclusions}
\label{sec:conclusions}
We present the first catalogue of UV-excess sources selected from a total of $7 \times 10^{6}$ stars 
in 211 square degrees of \UVEX data.
This catalogue contains 2\,170 unique UV-excess candidates with magnitudes between $14 < g < 22.5$. 
With the selection technique presented in Sect.\ \ref{sec:algorithm} we 
select on average $\sim$11 UV-excess candidates per square degree.
Cross-matching with other catalogues shows that the UV-excess catalogue consist of a mix of populations 
with only a small fraction that has been identified before.

\begin{figure*}
\centerline{\epsfig{figure=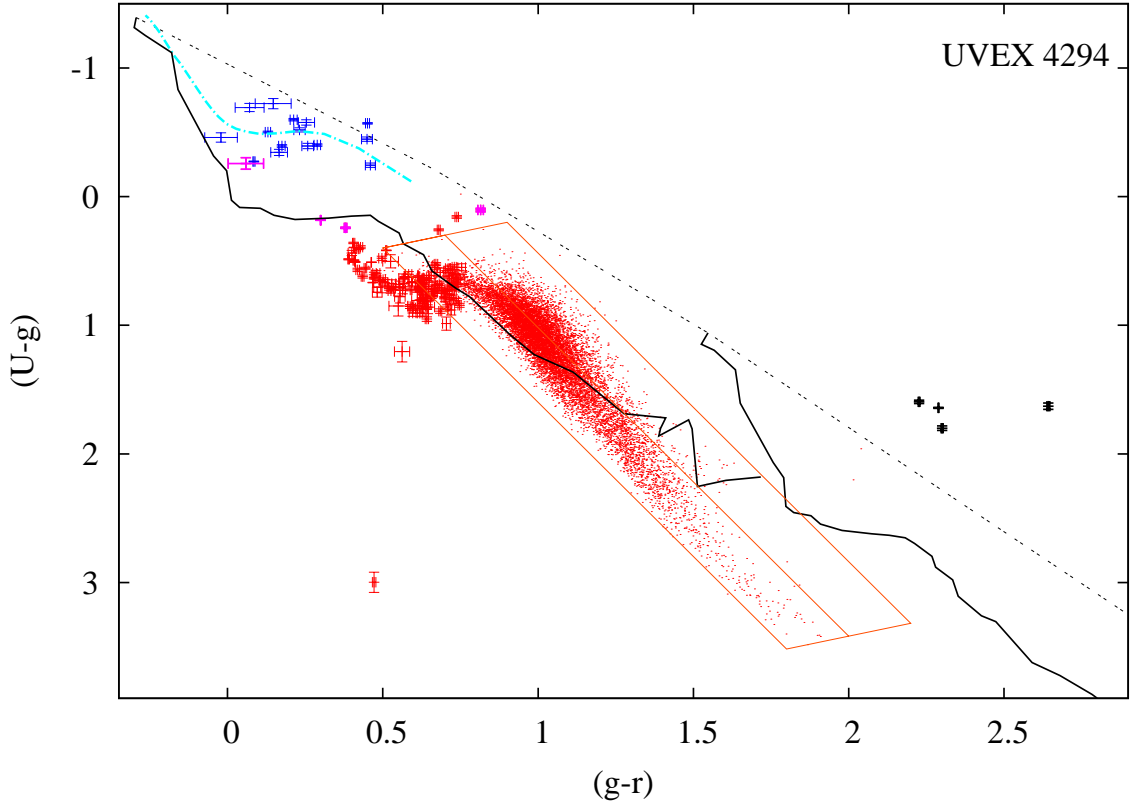,width=16cm,angle=0}}
\caption{$(U-g)$ vs. $(g-r)$ colour-colour diagram of field UVEX 4294. 
Data shown are all stellar and probably stellar detections (red dots) of field UVEX 4294 (l=43.88, b=4.67). 
The lines show the line fitted to the main locus (central line) and the 3$\sigma$ wide selection box.  
The blue, red and black data points are respectively the UV-excess, subdwarf and purple candidates selected from this field. 
The 4 magenta data points show subdwarfs selected in the blue corner of the colour-colour diagram outside the selection box but less 
than 0.4 magnitude from the blue edge in the colour-magnitude diagrams.
The black solid lines are the simulated colours of main-sequence stars with E(B-V)=0 and E(B-V)=2,  
the dashed lines are the unreddened DA white dwarf tracks (cyan) and the O5V-reddening line (black dashed).    
\label{fig:uvex4294ccd}}
\end{figure*}

\begin{figure}
\centerline{\epsfig{figure=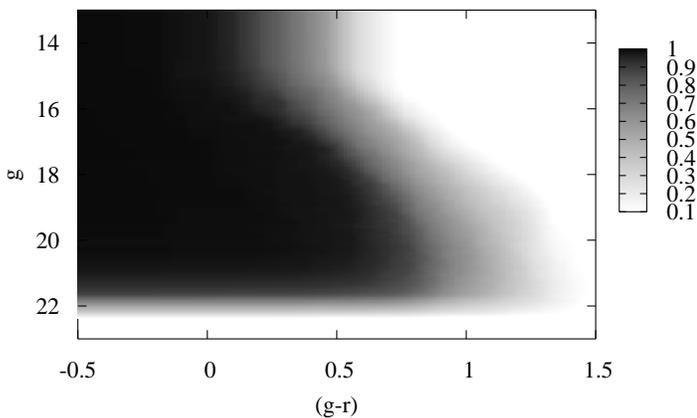,width=10.2cm,angle=0}}
\caption[]{Probability that a source will be picked-up by the selection algorithm of Sect.2 depending 
on the position of the source in the $g$ vs. $(g-r)$ colour-magnitude diagram.   
\label{fig:prob}}
\end{figure}

\begin{figure*}
\centerline{\psfig{figure=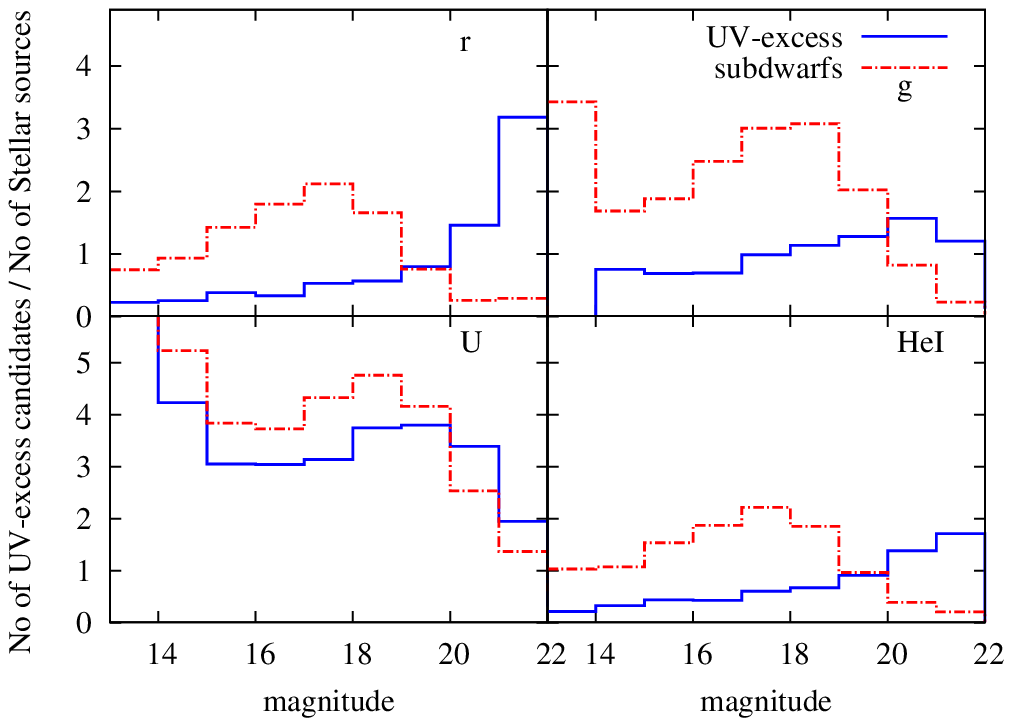,width=18cm,angle=0}}
\caption{$r$-, $g$-, $U$-, and $HeI$-band magnitude distributions of the UV-excess candidates (solid/blue)
and the `subdwarf' sample (dot-dashed/red). In the histograms the number of UV-excess candidates and 
subdwarfs is devided by the number of stellar sources in the magnitude bin and
normalized to one by the total number of UV-excess, subdwarf and stellar sources.
\label{fig:MagDistUVexcess}}
\end{figure*}

\subsection{Completeness and reliability of the UV-excess catalogue}
An illustration of the field-to-field selection algorithm was given in 
Figs.\ \ref{fig:uvex6162cmd} to \ \ref{fig:uvex6162cmd2}.
The technique works well, but is not perfect: main-sequence stars at the blue side of 
the main-sequence may still be selected when they are slightly scattered. 
In the colour-magnitude diagrams (see Figs.\ \ref{fig:uvex6162cmd} and \ \ref{fig:uvex6162cmd2}) the edge of the
selection line (left `ragged' solid line) may become rather `jumpy' when only a small number of
main-sequence objects is detected, leading to a loss of objects or the inclusion of too many objects at the brighter or fainter ends. 
For example in the $g$ vs. $(U-g)$ colour-magnitude diagram of Fig.\ \ref{fig:uvex6162cmd} there is 
one object at $g$$\sim$21 not selected as UV-excess candidate due to the blue edge of the last bin.  
The number of UV-excess candidates in the catalogue brighter than $g<16$ is 23.
Fig.\ \ref{fig:MagDistUVexcess} shows the magnitude distributions of the UV-excess sources. The number of selected UV-excess sources
increases for fainter magnitudes. The probability that a source with a particular $(g-r)$ colour 
and $g$-band magnitude will be picked-up in our 211 square degrees
is given in Fig.\ \ref{fig:prob}. For a given $(g-r)$ colour the 
probability for faint sources to be selected is larger than for bright objects.\\  

We checked the colour-magnitude and colour-colour plots during our selection method visually for $\sim$10 percent of the fields.
The field with the largest number of selected UV-excess candidates contains 15 sources.
The $(U-g)$ vs. $(g-r)$ colour-colour diagram of this field is shown in Fig.\ \ref{fig:uvex4294ccd}. 
In field \UVEX 4294 there are 19 sources selected in the blue corner of the $(U-g)$ vs. $(g-r)$ colour-colour diagram outside the selection box.  
Four of these sources (magenta in Fig.\ \ref{fig:uvex4294ccd}) are less than 0.4 magnitude 
from the blue edge in the colour-magnitude diagrams and are in the `subdwarf' sample.
Three of these `subdwarf' sources are probably early-type stars since they are too close to the main locus,
the fourth `subdwarf' is near the location of the UV-excess sources. 
The 15 UV-excess candidates in Fig.\ \ref{fig:uvex4294ccd} are more than 0.4 magnitude from the blue edge in the colour-magnitude diagrams 
and are on top of the unreddened white dwarf track.
In very sparse fields without blue or UV-excess sources the fit of the straight line 
in the $(U-g)$ vs. $(g-r)$ colour-colour diagram may be completely wrong, but this is no problem since 
these fields will not contribute new UV-excess objects to our catalogue anyway.\\

Guaranteeing that every object in the UV-excess catalogue is indeed a real UV-excess source is impossible, 
although first preliminary spectroscopic follow-up observations (Verbeek et al., 2011b, in prep.) 
show that at least 80$\%$ of the UV-excess candidates are indeed 
genuine UV-excess sources such as white dwarfs and white dwarf binaries.
As can be seen in Fig.\ \ref{fig:blueEdge} the tails of the $(g-r)$ distributions of the UV-excess sources and the subdwarf sample overlap. 
Both samples will contain some objects from the other category since they are separated at $\Delta (g-r)$=0.4 magnitudes from the blue edge.
In the colour-magnitude diagrams of Fig.\ \ref{fig:UVexcessSampleCMD} it can be seen that the UV-excess candidates and the subdwarf sample separate.
Still there is a fraction of scattered subdwarfs at the location of the UV-excess sources, see e.g. the fainter subdwarf sources with $(U-g)<0$ in the
$g$ vs. $(U-g)$ colour-magnitude diagram of Fig.\ \ref{fig:UVexcessSampleCMD}. The majority of these subdwarfs are selected only in $g$ vs. $(g-r)$ 
less than 0.4 magnitude from the blue edge.
Also at the location of the subdwarfs in Fig.\ \ref{fig:UVexcessSampleCMD} there are $\sim$100 scattered 
UV-excess sources, selected in several different ways.
The field-to-field selection algorithm can be influenced in crowded regions and regions where extinction varies strongly.
The selection algorithm can still be used when there are magnitude offsets in the global photometric calibration.
Only when there is a large offset in $(g-r)$ the demand that UV-excess sources must be bluer than $(g-r)$=1.25 will cause problems.
That is why 33 fields with large photometric shifts were removed during the selection.
The aim to have the highest possible percentage real UV-excess sources in our sample 
forces us to use a conservative selection method.\\ 

\subsection{The two additional subdwarf and purple samples}
Two additional sub-samples of 9\,872 objects labelled as `subdwarf' candidates and 
803 objects labelled as `purple' sources are selected from the 211 square degrees of \UVEX data.
The subdwarf sample is slightly bluer than the main-sequence and contains a 
mix of metal-poor stars and lightly reddened main-sequence stars.
Fig.\ \ref{fig:uvex4294ccd} shows the colour-colour diagram of field \UVEX 4294 where
most `subdwarf' sources, selected in the $g$ vs. $(g-r)$ colour-magnitude diagram, 
overlap with the earliest main-sequence stars in this field.
Fig.\ \ref{fig:MagDistUVexcess} shows the number of subdwarfs divided by the number of
stellar sources per magnitude bin. There is a bump in the $g$-band and $U$-band magnitude distributions for the bright subdwarfs.
The bright subdwarfs are selected in the $g$ vs. $(g-r)$ colour-magnitude diagram
and in the $g$ vs. $(U-g)$ colour-magnitude diagram but not in the $(U-g)$ vs. $(g-r)$ colour-colour diagram.
This is because the first bins of the blue edge in Fig.\ \ref{fig:uvex6162cmd} are usually larger, while the sources 
of the main locus change over $(g-r)$ or $(U-g)$ in these first bins.
In field \UVEX 4294 there are 9 sources at the purple side more than 3 $\sigma$ outside the selection box. 
Only the 4 purple sources that are above the O5V-reddening line are selected in the `purple' sample 
since the other sources might be scattered reddened main-sequence stars. Only when there are 
large magnitude offsets in $(g-r)$ or $(U-g)$ this demand might cause problems. 
Fields with obvious magnitude offsets in $(g-r)$, $(U-g)$ or in $(HeI-r)$ were removed in Sect.\ \ref{sec:selection}.
In the $(U-g)$ vs. $(g-r)$ colour-colour diagram
it is not possible to use the reddening vector to fit a straight line to the main locus since each field will
consist of a mix of populations at different distances subjected to different reddenings.\\

The cross-matching with other catalogues and databases (Deacon, Corradi, Viironen, Witham, 2MASS, UKIDSS and Simbad)
shows that $\sim$1 percent of the UV-excess sources were previously known objects and $\sim$77 percent 
were previously detected by \IPHAS. The vast majority of objects are not classified yet, 
so \UVEX uncovers a large new sample of white dwarfs, subdwarfs, 
Cataclysmic Variables, symbiotic stars, planetary nebulae 
and other post-common-envelope objects.\\

\section*{Acknowledgement}
This paper makes use of data collected at the Isaac Newton Telescope, operated on the island of La Palma by the Isaac Newton Group in the
Spanish Observatorio del Roque de los Muchachos of the Inst\'{\i}tuto de Astrof\'{\i}sica de Canarias.
The observations were processed by the Cambridge Astronomy Survey Unit
(CASU) at the Institute of Astronomy, University of Cambridge.
This work is based in part on data obtained as part of the UKIRT Infrared Deep Sky Survey (UKIDSS). 
We acknowledge the use of data products from Two-Micron All-Sky Survey (2MASS), which is a joint project of the University of
Massachusetts and the Infrared Processing and AnalysisCenter/California Institute of Technology (funded by the National
Aeronautics and Space Administration and National Science Foundation of the USA).
This research has made use of the Simbad database and the VizieR catalogue access 
tool, operated at CDS, Strasbourg, France.
KV is supported by a NWO-EW grant 614.000.601 to PJG. 
The authors would like to thank Detlev Koester and Pierre Bergeron for
making available their white dwarf models on which a significant part
of the colour simulations in this paper are based.
The authors want to thank all \UVEX observers for going to La Palma.\\

\label{lastpage}

\newpage

\appendix

\section{Catalogue of UV-excess candidates}
\label{app:uvexcesscatalogue}
The complete UV-excess catalogue, and the two additional `subdwarf' and `purple' samples, selected from the first 211 square degrees of \UVEX data 
can be obtained at the \UVEX website\footnote{http://www.uvexsurvey.org} or via VizieR\footnote{http://vizier.u-strasbg.fr/viz-bin/VizieR}.
The UV-excess catalogue, the `subdwarf' and `purple' samples contain 50 columns:\\ 

\noindent
1--6: RA and dec.\\
7: \UVEX field.\\
8--19: $r_{av}$, $\Delta r$, morp.class $r$, $g_{av}$, $\Delta g$, morp.class $g$, $U_{av}$, $\Delta U$, morp.class $U$, $He_{av}$, $\Delta He$, morp.class $He$.\\ 
20: \UVEX selection flag.\\
21: $(U-g)_{BlueEdge}$ - $(U-g)_{source}$ - $2\sigma_{(U-g)}$; the difference in $(U-g)$ between source and the blue edge in 
the $g$ vs. $(U-g)$ colour-magnitude diagram, inclusive the photometric error of the source.\\
22: $(g-r)_{BlueEdge}$ - $(g-r)_{source}$ - $2\sigma_{(g-r)}$; the difference in $(g-r)$ between source and the blue edge in 
the $g$ vs. $(g-r)$ colour-magnitude diagram, inclusive the photometric error of the source.\\
23--34: Original magnitudes, magnitude errors and morphology classes from the direct field: 
$r$, $\Delta r$, morp.class $r$, $g$, $\Delta g$, morp.class $g$, $U$, $\Delta U$, morp.class $U$, $He$, $\Delta He$, morp.class $He$.\\
35--46: Original magnitudes, magnitude errors and morphology classes from offset field: 
$r$, $\Delta r$, morp.class $r$, $g$, $\Delta g$, morp.class $g$, $U$, $\Delta U$, morp.class $U$, $He$, $\Delta He$, morp.class $He$.\\
47--50: Average $r_{av.field}$, $g_{av.field}$, $U_{av.field}$, $HeI_{av.field}$ magnitude of the total field.\\

-- Columns 8, 11, 14 and 17 contain averaged magnitudes ($r_{av}$, $g_{av}$, $U_{av}$ and $He_{av}$) over direct field and offset field. 
These magnitudes, corrected for magnitude shifts between field and offset field, are calculated in a few steps, e.g. for $r_{av}$:\\
i) first $(r_{direct}-r_{offset})$ is calculated for each coordinate match between a field and offset field,\\  
ii) from this $(r_{direct}-r_{offset})$ distribution the first 25$\%$ and last 25$\%$ is removed,\\
iii) from the central 50$\%$ the mean magnitude shift ($r_{shift}$) between direct field and offset field is calculated,\\
iv) the magnitudes of the offset field are corrected for magnitude shifts by $r_{offsetNEW}=r_{offset}-|r_{shift}|$,\\
v) per source the magnitude difference of direct field and offset field is checked: do the magnitudes match within the error bars? Here
an extra 0.05 mag is taken into account to avoid that too much sources are removed.\\ 
vi) the averaged magnitude $r_{av}$ is now simply calculated by: $r_{av}$=$(r_{direct}+r_{offsetNEW})/2$. \\
 
-- Columns 10, 13, 16 and 19 are ``averaged'' morphology class flags of the direct field and offset field: the best flag of the two. 
The possible morphology classes are: -1 (stellar), -2 (probably stellar) and 0 (noise-like).\\

-- Column 20 is the \UVEX selection flag (1, 2, 4, 32, 64, 512, 1024 or combinations) and gives information about how the source was selected:\\
1) outlier on the blue side of the main-sequence in the $(U-g)$ vs. $(g-r)$ colour-colour diagram.\\
2) outlier on the blue side of the main-sequence in the $g$ vs. $(U-g)$ colour-magnitude diagram.\\
4) outlier on the blue side of the main-sequence in the $g$ vs. $(g-r)$ colour magnitude diagram.\\
32) outlier on the purple side of the main-sequence in the $(U-g)$ vs. $(g-r)$ colour-colour diagram.\\
64) outlier on the non-purple side of the main-sequence in the $(U-g)$ vs. $(g-r)$ colour-colour diagram.\\
512) outlier on the blue side of the main-seq in the $g$ vs. $(U-g)$ colour-magnitude diagram that lies more than 0.4 mag away from the main-sequence blue edge.\\
1024) outlier on the blue side of the main-seq in the $g$ vs. $(g-r)$ colour-magnitude diagram that lies more than 0.4 mag away from the main-sequence blue edge.\\


\section{Synthetic colours with the effect of the $U$-band filter ``red leak''.}
\label{app:redleakuvexcolours}

\begin{figure}
\centerline{\epsfig{figure=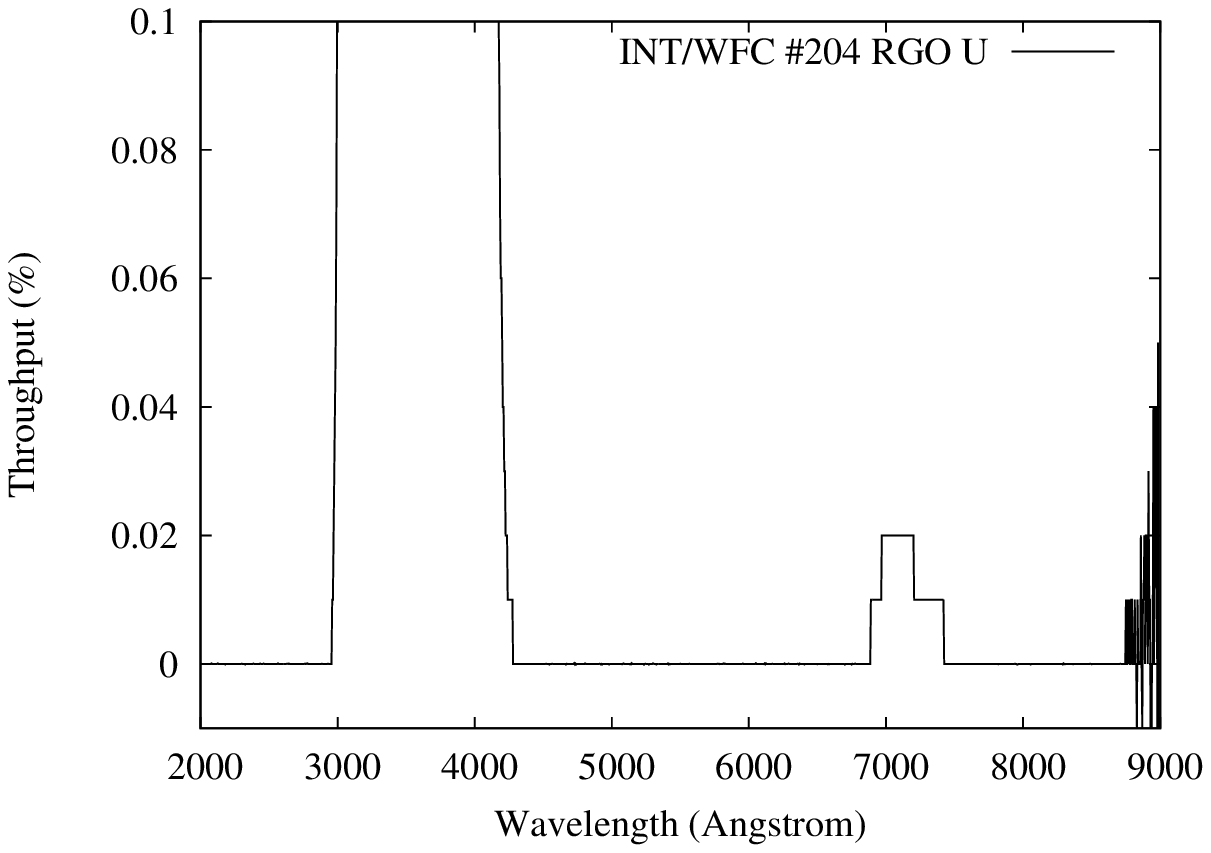,width=8.6cm}}
\caption[]{Throughput of the INT/WFC RGO U-band filter used for UVEX with the red leak near 7050\AA. 
\label{fig:redleakfilter}}
\end{figure}

\begin{figure}
\centerline{\epsfig{figure=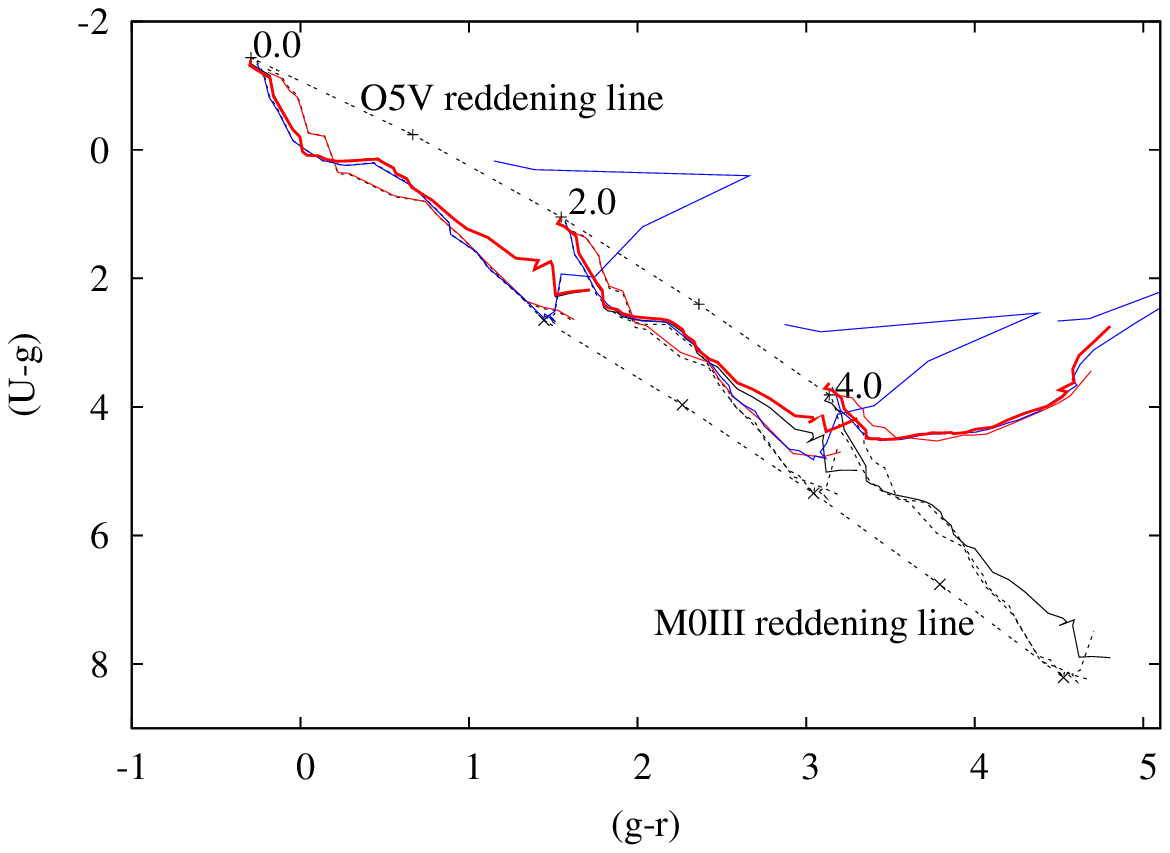,width=8.6cm}}
\caption[]{The synthetic colours of main-sequence stars, giants 
and super giants including the effect of the $U$-band filter red leak plotted in the $(U-g)$ vs. $(g-r)$ 
colour-colour diagram. The colour-colour diagram without the effect of the $U$-band red leak was already shown in G09 (Fig.6). 
In this graph the black lines are the old G09 colours, the overplotted lines are the new synthetic colours of 
main-sequence stars (thick red), giants (blue) and super giants (thin red) for E(B--V)=0.0, 2.0 and 4.0 using the reddening law of Cardelli et al. (1989). 
The upper and lower envelope curves (dashed) are the old O5V-reddening curve and the M0III-reddening curve of G09.
Late M-giants from M7III to M10III, not shown in G09, were add.
\label{fig:redleakcolours}}
\end{figure}

The \UVEX $U$-band filter has a small ``red leak'' near 7050\AA , just large enough 
to make very red sources appear slightly brighter in $U$ than they should be. 
This red leak, shown in Fig.\ \ref{fig:redleakfilter}, has more effect on late 
type main-sequence stars than on early type main-sequence stars and contributes
stronger for highly reddened stars. 
For unreddened M-dwarfs the effect of this red leak is in the range of 0.01--0.03 magnitudes in $(U-g)$.
For M-dwarfs reddened by $E(B-V)=1$ the effect is in the range of 0.1--0.3 magnitudes in $(U-g)$.
This effect for different $E(B-V)$ is shown in Fig.\ \ref{fig:redleakcolours} together 
with the synthetic \UVEX colours of G09.
Also with the effect of the $U$-band filter red leak all late type main-sequence stars and giants with reddening smaller than 
$E(B-V)=2$ are still below the O5V-reddening curve. We do not expect to pick up 
very late type main-sequence stars with reddening larger than 
$E(B-V)=2$ because they would be too faint. Late type giants with reddening larger than $E(B-V)=2$ 
might be picked up by the selection algorithm since they are intrinsically brighter.
The purple sample will contain a significant number of late-type M-giants since the 
synthetic colours of M-giants later than type M5III 
are clearly in the region above the O5V-reddening line, see Fig.\ \ref{fig:redleakcolours}.
The effect of the red leak was not included in the tables of G09 
since the existence of the leak was not known yet at that time.
Tables with new synthetic main-sequence, giant, supergiant and 
white dwarf colours for the \UVEX filter system including this effect 
of the $U$-band red leak, and including reddening, are given here. 
Only the $(U-g)$ colours are different compared to G09, for completeness the other \UVEX
colours are also given here.\\

\oddsidemargin=-2cm
\evensidemargin=-2cm

\appendix

\oddsidemargin=-2cm
\evensidemargin=-2cm


\renewcommand*\thetable{\Alph{section}B\arabic{table}}

\begin{table*}
\caption[]{\UVEX colour indices $(U-g)$, $(g-r)$, $(HeI-r)$ for Pickles main-sequence stars including
  reddening. \label{tab:mainsequence} }
\small
\noindent
\begin{tabular}{p{3mm}@{\ \ \ }p{9mm}@{\ \ \ }p{9mm}@{\ \ \ }p{10mm}@{\ \ \ }p{9mm}@{\ \ \ }p{8mm}@{\ \ \ }p{10mm}@{\ \ \ }p{8mm}@{\ \ \ }p{8mm}@{\ \ \ }p{10mm}@{\ \ \ }p{8mm}@{\ \ \ }p{8mm}@{\ \ \ }p{10mm}@{\ \ \ }p{8mm}@{\ \ \ }p{8mm}@{\ \ \ }p{10mm}}
\hline 
\multicolumn{1}{l}{Spec.}&
\multicolumn{3}{l}{$E(B-V)=0.0$}&
\multicolumn{3}{l}{$E(B-V)=1.0$}&
\multicolumn{3}{l}{$E(B-V)=2.0$}&
\multicolumn{3}{l}{$E(B-V)=3.0$}&
\multicolumn{3}{l}{$E(B-V)=4.0$}
\\
type & $U$$-$$g$ & $g$$-$$r$ & $He${\sc i}$-$$r$ & $U$$-$$g$ & $g$$-$$r$ &
$He${\sc i}$-$$r$ & $U$$-$$g$ & $g$$-$$r$ & $He${\sc i}$-$$r$&
  $U$$-$$g$ & $g$$-$$r$ & $He${\sc i}$-$$r$& $U$$-$$g$ & $g$$-$$r$ &
  $He${\sc i}$-$$r$ \\ \hline
O5V & --1.393 & --0.294 &  0.006  & --0.209 &  0.667 &  0.149  &   1.059 &  1.547 &  0.326   &2.385 &  2.364 &  0.538 &  3.632 &  3.137 &  0.784     \\ 
O9V & --1.314 & --0.301 &  0.008  & --0.125 &  0.652 &  0.151  &   1.148 &  1.525 &  0.329   &2.475 &  2.338 &  0.541 &  3.712 &  3.111 &  0.788     \\
B0V & --1.254 & --0.265 &  0.006  & --0.072 &  0.691 &  0.148  &   1.194 &  1.567 &  0.326   &2.515 &  2.381 &  0.538 &  3.738 &  3.152 &  0.784     \\
B1V & --1.119 & --0.180 &  0.001  &  0.072  &  0.767 &  0.146   &  1.346 &  1.635 &  0.326   &2.670 &  2.442 &  0.541 &  3.864 &  3.209 &  0.788    \\
B3V & --0.833 & --0.160 &  0.004  &  0.346  &  0.785 &  0.151   &  1.606 &  1.651 &  0.332   &2.907 &  2.458 &  0.548 &  4.031 &  3.226 &  0.798    \\
B8V & --0.316 & --0.045 &  0.005  &  0.835  &  0.895 &  0.153   &  2.066 &  1.757 &  0.337   &3.323 &  2.561 &  0.555 &  4.294 &  3.326 &  0.806    \\
B9V & --0.202 & --0.003 &  0.009  &  0.952  &  0.932 &  0.160   &  2.184 &  1.791 &  0.345   &3.436 &  2.592 &  0.564 &  4.364 &  3.355 &  0.816    \\
A0V &  0.028 &  0.013 &  0.002  &    1.180  &  0.943 &  0.153  &   2.408 &  1.798 &  0.338   &3.643 &  2.596 &  0.558 &  4.486 &  3.357 &  0.811 \\
A2V &  0.084 &  0.039 & --0.002 &    1.232  &  0.969 &  0.149  &   2.456 &  1.823 &  0.336   &3.682 &  2.621 &  0.557 &  4.496 &  3.383 &  0.812  \\
A3V &  0.090 &  0.106 &  0.011  &    1.248  &  1.029 &  0.164  &   2.481 &  1.878 &  0.352   &3.709 &  2.672 &  0.575 &  4.494 &  3.431 &  0.831 \\
A5V &  0.146 &  0.150 &  0.017  &    1.309  &  1.066 &  0.171  &   2.545 &  1.909 &  0.361   &3.768 &  2.698 &  0.585 &  4.510 &  3.454 &  0.842 \\
A7V &  0.178 &  0.217 &  0.024  &    1.351  &  1.134 &  0.182  &   2.595 &  1.978 &  0.374   &3.814 &  2.769 &  0.600 &  4.498 &  3.526 &  0.859 \\
F0V &  0.167 &  0.325 &  0.046  &    1.360  &  1.235 &  0.208  &   2.621 &  2.074 &  0.405   &3.839 &  2.862 &  0.637 &  4.459 &  3.618 &  0.901 \\
F2V &  0.151 &  0.399 &  0.070  &    1.359  &  1.299 &  0.235  &   2.632 &  2.131 &  0.435   &3.851 &  2.914 &  0.669 &  4.429 &  3.667 &  0.935 \\
F5V &  0.145 &  0.460 &  0.068  &    1.367  &  1.355 &  0.234  &   2.652 &  2.182 &  0.436   &3.874 &  2.962 &  0.670 &  4.423 &  3.713 &  0.937 \\
F6V &  0.192 &  0.486 &  0.074  &    1.413  &  1.382 &  0.242  &   2.696 &  2.212 &  0.445   &3.905 &  2.994 &  0.682 &  4.402 &  3.747 &  0.951 \\
F8V &  0.282 &  0.552 &  0.086  &    1.510  &  1.443 &  0.256  &   2.796 &  2.268 &  0.462   &3.986 &  3.047 &  0.701 &  4.406 &  3.798 &  0.971 \\
G0V &  0.367 &  0.567 &  0.096  &    1.596  &  1.455 &  0.268  &   2.881 &  2.280 &  0.474   &4.052 &  3.060 &  0.714 &  4.405 &  3.811 &  0.985 \\
G2V &  0.451 &  0.631 &  0.100  &    1.690  &  1.514 &  0.273  &   2.980 &  2.335 &  0.481   &4.129 &  3.112 &  0.722 &  4.399 &  3.862 &  0.995 \\
G5V &  0.579 &  0.658 &  0.106  &    1.819  &  1.536 &  0.280  &   3.107 &  2.353 &  0.489   &4.227 &  3.127 &  0.732 &  4.417 &  3.875 &  1.005 \\
G8V &  0.710 &  0.734 &  0.130  &    1.965  &  1.610 &  0.307  &   3.258 &  2.427 &  0.519   &4.333 &  3.202 &  0.764 &  4.395 &  3.951 &  1.040 \\
K0V &  0.778 &  0.779 &  0.150  &    2.023  &  1.656 &  0.331  &   3.304 &  2.474 &  0.545   &4.345 &  3.250 &  0.793 &  4.350 &  4.000 &  1.071 \\
K2V &  1.082 &  0.913 &  0.164  &    2.345  &  1.780 &  0.346  &   3.622 &  2.591 &  0.563   &4.538 &  3.361 &  0.813 &  4.315 &  4.107 &  1.093 \\
K3V &  1.228 &  0.988 &  0.201  &    2.474  &  1.864 &  0.387  &   3.724 &  2.681 &  0.607   &4.541 &  3.454 &  0.859 &  4.199 &  4.201 &  1.143 \\
K4V &  1.364 &  1.113 &  0.235  &    2.628  &  1.976 &  0.425  &   3.870 &  2.784 &  0.649   &4.572 &  3.553 &  0.906 &  4.100 &  4.297 &  1.193 \\
K5V &  1.687 &  1.277 &  0.252  &    2.947  &  2.133 &  0.443  &   4.138 &  2.935 &  0.667   &4.622 &  3.698 &  0.923 &  3.985 &  4.436 &  1.209 \\
K7V &  1.720 &  1.411 &  0.328  &    2.992  &  2.250 &  0.526  &   4.163 &  3.038 &  0.758   &4.536 &  3.791 &  1.021 &  3.837 &  4.522 &  1.313 \\
M0V &  1.802 &  1.394 &  0.378  &    3.063  &  2.245 &  0.580  &   4.205 &  3.044 &  0.816   &4.505 &  3.806 &  1.083 &  3.767 &  4.545 &  1.379 \\
M1V &  1.861 &  1.389 &  0.425  &    3.119  &  2.227 &  0.628  &   4.243 &  3.015 &  0.865   &4.495 &  3.768 &  1.133 &  3.736 &  4.501 &  1.430 \\
M2V &  1.736 &  1.485 &  0.491  &    3.005  &  2.320 &  0.702  &   4.134 &  3.105 &  0.946   &4.379 &  3.857 &  1.220 &  3.619 &  4.590 &  1.523 \\
M3V &  1.806 &  1.496 &  0.626  &    3.077  &  2.320 &  0.848  &   4.148 &  3.098 &  1.102   &4.238 &  3.848 &  1.386 &  3.416 &  4.581 &  1.697 \\
M4V &  2.254 &  1.513 &  0.742  &    3.506  &  2.336 &  0.977  &   4.385 &  3.119 &  1.244   &4.127 &  3.876 &  1.542 &  3.199 &  4.619 &  1.866 \\
M5V &  2.206 &  1.605 &  0.769  &    3.466  &  2.423 &  1.006  &   4.299 &  3.202 &  1.275   &3.964 &  3.956 &  1.573 &  3.015 &  4.696 &  1.896 \\
M6V &  2.179 &  1.717 &  0.784  &    3.447  &  2.526 &  1.038  &   4.179 &  3.302 &  1.323   &3.720 &  4.059 &  1.635 &  2.746 &  4.803 &  1.972  \\ \hline   
\end{tabular}                                                                                
\end{table*}

\newpage

\begin{table*}
\label{tab:giants}
\caption[]{\UVEX colour indices $(U-g)$, $(g-r)$, $(HeI-r)$ for Pickles Giants including reddening.}
\small
\begin{tabular}{p{10mm}@{\ \ \ }p{9mm}@{\ \ \ }p{9mm}@{\ \ \ }p{10mm}@{\ \ \ }p{9mm}@{\ \ \ }p{7mm}@{\ \ \ }p{10mm}@{\ \ \ }p{7mm}@{\ \ \ }p{7mm}@{\ \ \ }p{10mm}@{\ \ \ }p{7mm}@{\ \ \ }p{7mm}@{\ \ \ }p{10mm}@{\ \ \ }p{7mm}@{\ \ \ }p{7mm}@{\ \ \ }p{10mm}}
\hline 
\multicolumn{1}{l}{Spec.}&
\multicolumn{3}{l}{$E(B-V)$=0.0}&
\multicolumn{3}{l}{$E(B-V)$=1.0}&
\multicolumn{3}{l}{$E(B-V)$=2.0}&
\multicolumn{3}{l}{$E(B-V)$=3.0}&
\multicolumn{3}{l}{$E(B-V)$=4.0}\\
type    & $U$$-$$g$ & $g$$-$$r$ & $He${\sc i}$-$$r$ &$U$$-$$g$ &
$g$$-$$r$ & $He${\sc i}$-$$r$ & $U$$-$$g$ & $g$$-$$r$ & $He${\sc
  i}$-$$r$& $U$$-$$g$ & $g$$-$$r$ & $He${\sc i}$-$$r$& $U$$-$$g$ &
$g$$-$$r$ & $He${\sc i}$-$$r$  \\ \hline
O8III   & --1.341 & --0.259 &  0.001  & --0.152 &  0.696 &  0.143  &  1.122 &  1.571 &  0.320&   2.450 &  2.383 &  0.532  &  3.689  &  3.154 &  0.777  \\
B1-2III & --1.133 & --0.215 &  0.016  &  0.053  &  0.733 &  0.160  &  1.322 &  1.603 &  0.340&   2.642 &  2.412 &  0.554  &  3.841  &  3.181 &  0.801  \\
B3III   & --0.820 & --0.185 &  0.032  &  0.362  &  0.759 &  0.179  &  1.625 &  1.627 &  0.361&   2.931 &  2.435 &  0.578 &   4.057  &  3.204 &  0.828	 \\
B5III   & --0.594 & --0.132 &  0.010  &  0.577  &  0.812 &  0.158  &  1.828 &  1.679 &  0.341&   3.114 &  2.487 &  0.559  &  4.175  &  3.256 &  0.810	 \\
B9III   & --0.143 & --0.045 & --0.002 &  0.999  &  0.895 &  0.147  &  2.220 &  1.757 &  0.330&   3.461 &  2.561 &  0.548  &  4.385  &  3.326 &  0.799	 \\
A0III   &  0.016  &  0.040 & --0.003  &  1.172  &  0.968 &  0.147  &  2.405 &  1.821 &  0.333 &  3.642  &  2.618 &  0.552  & 4.483  &  3.377 &  0.805	  \\
A3III   &  0.168  &  0.136 & --0.008  &  1.324  &  1.054 &  0.145  &  2.553 &  1.899 &  0.334 &  3.772  &  2.689 &  0.556  & 4.526  &  3.445 &  0.812	  \\
A5III   &  0.185  &  0.172 &  0.031  &   1.351 &  1.088 &  0.186  &   2.590 &  1.932 &  0.377  & 3.807  &  2.722 &  0.603  & 4.511  &  3.480 &  0.861   \\
A7III   &  0.230  &  0.234 &  0.025  &   1.398 &  1.150 &  0.183  &   2.638 &  1.994 &  0.376  & 3.849  &  2.784 &  0.602  & 4.511  &  3.541 &  0.862   \\
F0III   &  0.242  &  0.276 &  0.038  &   1.418 &  1.186 &  0.198  &   2.663 &  2.026 &  0.393  & 3.872  &  2.815 &  0.622  & 4.503  &  3.571 &  0.883	\\
F2III   &  0.206  &  0.437 &  0.060  &   1.408 &  1.337 &  0.226  &   2.677 &  2.169 &  0.427  & 3.884  &  2.952 &  0.662  & 4.427  &  3.704 &  0.929	\\
F5III   &  0.241  &  0.451 &  0.074  &   1.446 &  1.352 &  0.242  &   2.715 &  2.186 &  0.444  & 3.915  &  2.971 &  0.680  & 4.424  &  3.726 &  0.948	\\
G0III   &  0.564  &  0.663 &  0.106  &   1.801 &  1.544 &  0.281  &   3.087 &  2.364 &  0.490  & 4.209  &  3.140 &  0.732  & 4.411  &  3.888 &  1.006	\\
G5III   &  0.978  &  0.822 &  0.127  &   2.242 &  1.686 &  0.307  &   3.528 &  2.493 &  0.522  & 4.511  &  3.261 &  0.770  & 4.392  &  4.007 &  1.048	\\
G8III   &  1.136  &  0.883 &  0.133  &   2.396 &  1.740 &  0.314  &   3.669 &  2.543 &  0.529  & 4.583  &  3.307 &  0.777  & 4.365  &  4.049 &  1.055	\\
K0III   &  1.313  &  0.891 &  0.162  &   2.569 &  1.753 &  0.346  &   3.828 &  2.559 &  0.564  & 4.669  &  3.328 &  0.815  & 4.356  &  4.074 &  1.096	\\
K1III   &  1.458  &  0.972 &  0.162  &   2.721 &  1.826 &  0.346  &   3.967 &  2.628 &  0.565  & 4.720  &  3.393 &  0.816  & 4.304  &  4.137 &  1.098	\\
K2III   &  1.589  &  1.047 &  0.186  &   2.842 &  1.897 &  0.374  &   4.066 &  2.696 &  0.597  & 4.736  &  3.458 &  0.852  & 4.248  &  4.200 &  1.137	\\
K3III   &  1.842  &  1.118 &  0.191  &   3.103 &  1.960 &  0.380  &   4.300 &  2.753 &  0.603  & 4.818  &  3.513 &  0.859  & 4.208  &  4.252 &  1.145	\\
K4III   &  2.259  &  1.303 &  0.198  &   3.539 &  2.123 &  0.391  &   4.657 &  2.900 &  0.619  & 4.856  &  3.647 &  0.878  & 4.073  &  4.377 &  1.168	\\
K5III   &  2.364  &  1.345 &  0.252  &   3.621 &  2.175 &  0.452  &   4.681 &  2.959 &  0.684  & 4.770  &  3.712 &  0.949  & 3.944  &  4.448 &  1.244	\\
M0III   &  2.620  &  1.444 &  0.327  &   3.867 &  2.266 &  0.532  &   4.823 &  3.044 &  0.770  & 4.703  &  3.793 &  1.041  & 3.813  &  4.527 &  1.341	\\
M1III   &  2.554  &  1.449 &  0.313  &   3.795 &  2.272 &  0.518  &   4.764 &  3.050 &  0.757  & 4.671  &  3.800 &  1.028  & 3.790  &  4.533 &  1.328	\\
M2III   &  2.675  &  1.513 &  0.385  &   3.913 &  2.337 &  0.597  &   4.803 &  3.117 &  0.844  & 4.578  &  3.870 &  1.122  & 3.663  &  4.607 &  1.429	\\
M3III   &  2.572  &  1.477 &  0.470  &   3.808 &  2.302 &  0.688  &   4.704 &  3.084 &  0.939  & 4.489  &  3.838 &  1.221  & 3.573  &  4.578 &  1.532	\\
M4III   &  2.512  &  1.507 &  0.629  &   3.734 &  2.335 &  0.858  &   4.568 &  3.120 &  1.121  & 4.272  &  3.880 &  1.415  & 3.338  &  4.625 &  1.736   \\
M5III   &  1.930  &  1.547 &  0.757  &   3.161 &  2.390 &  0.997  &   4.115 &  3.186 &  1.270  & 4.001  &  3.954 &  1.573  & 3.112  &  4.707 &  1.903   \\   
M6III   &  1.981  &  1.745 &  0.944  &   3.181 &  2.599 &  1.203  &   3.979 &  3.405 &  1.494  & 3.648  &  4.183 &  1.814  & 2.705  &  4.947 &  2.159	\\
M7III   &  1.199  &  2.029 &  1.203  &   2.403 &  2.905 &  1.484  &   3.287 &  3.724 &  1.795  & 3.080  &  4.512 &  2.132  & 2.169  &  5.283 &  2.494	\\
M8III   &  0.403  &  2.663 &  1.705  &   1.655 &  3.551 &  2.020  &   2.541 &  4.377 &  2.362  & 2.291  &  5.169 &  2.725  & 1.378  &  5.943 &  3.108	\\
M9III   &  0.309  &  1.389 &  1.344  &   1.615 &  2.257 &  1.629  &   2.831 &  3.086 &  1.946  & 3.287  &  3.890 &  2.289  & 2.626  &  4.681 &  2.656	\\
M10III  &  0.172  &  1.149 &  1.543  &   1.479 &  2.031 &  1.840  &   2.719 &  2.872 &  2.167  & 3.269  &  3.689 &  2.522  & 2.665  &  4.492 &  2.899	\\ \hline
\end{tabular} 
\end{table*}

\newpage

\begin{table*}
\caption[]{\UVEX/\IPHAS colour indices $(U-g)$, $(g-r)$, $(HeI-r)$ for Pickles Supergiants including reddening.}
\begin{tabular}{p{3mm}@{\ \ \ }p{9mm}@{\ \ \ }p{9mm}@{\ \ \ }p{10mm}@{\ \ \ }p{9mm}@{\ \ \ }p{9mm}@{\ \ \ }p{10mm}@{\ \ \ }p{8mm}@{\ \ \ }p{8mm}@{\ \ \ }p{10mm}@{\ \ \ }p{8mm}@{\ \ \ }p{8mm}@{\ \ \ }p{10mm}@{\ \ \ }p{8mm}@{\ \ \ }p{8mm}@{\ \ \ }p{10mm}}
\hline 
\multicolumn{1}{l}{Spec.}&
\multicolumn{3}{l}{$E(B-V)$=0.0}&
\multicolumn{3}{l}{$E(B-V)$=1.0}&
\multicolumn{3}{l}{$E(B-V)$=2.0}&
\multicolumn{3}{l}{$E(B-V)$=3.0}&
\multicolumn{3}{l}{$E(B-V)$=4.0}\\
type & $U$$-$$g$ & $g$$-$$r$ & $He${\sc i}$-$$r$ &$U$$-$$g$ &
$g$$-$$r$ & $He${\sc i}$-$$r$ & $U$$-$$g$ & $g$$-$$r$ &$He${\sc
  i}$-$$r$ & $U$$-$$g$ & $g$$-$$r$ & $He${\sc i}$-$$r$ & $U$$-$$g$ &
$g$$-$$r$ & $He${\sc i}$-$$r$\\ \hline
B0I & --1.210 & --0.193 &  0.040 & --0.005 &  0.748 &  0.186 &  1.280  &1.613 & 0.368&    2.612 &  2.420 &  0.585 &   3.817  &  3.189 &  0.834    \\
B1I & --1.121 & --0.118 &  0.034 &  0.082 &  0.825 &  0.186 &   1.364  &1.692 & 0.374&    2.689 &  2.502 &  0.596 &   3.857  &  3.274 &  0.851    \\
B3I & --0.918 & --0.064 &  0.041 &  0.270 &  0.878 &  0.190 &   1.540  &1.741 & 0.374&    2.851 &  2.545 &  0.593 &   3.977  &  3.309 &  0.845    \\
B5I & --0.821 & --0.019 &  0.052 &  0.375 &  0.913 &  0.206 &   1.649  &1.771 & 0.395 &   2.957 &  2.573 &  0.619 &   4.040  &  3.339 &  0.875   \\
B8I & --0.636 &  0.005 &  0.067 &   0.552 &  0.931 &  0.221  &  1.818  &1.783 & 0.410 &   3.113 &  2.581 &  0.634 &   4.152  &  3.344 &  0.890  \\
A0I & --0.260 &  0.047 &  0.051 &   0.890 &  0.977 &  0.206  &  2.120  &1.832 &  0.397&   3.369 &  2.632 &  0.621 &   4.292  &  3.397 &  0.879   \\
A2I & --0.211 &  0.143 &  0.046 &   0.954 &  1.064 &  0.202  &  2.197  &1.912 &  0.394&   3.449 &  2.706 &  0.620 &   4.322  &  3.465 &  0.878   \\
F0I &  0.353  &  0.224 &  0.048 &   1.488 &  1.141 &  0.208 &	2.696  &1.986 &  0.403  & 3.874  &  2.780 &  0.633 &  4.496   &  3.541 &  0.896  \\ 
F5I &  0.363  &  0.288 &  0.069 &   1.511 &  1.204 &  0.234 &	2.731  &2.050 &  0.433  & 3.908  &  2.846 &  0.666 &  4.481   &  3.609 &  0.931  \\ 
F8I &  0.715  &  0.558 &  0.077 &   1.906 &  1.438 &  0.244 &	3.152  &2.255 &  0.446  & 4.264  &  3.029 &  0.681 &  4.531   &  3.776 &  0.948  \\ 
G0I &  0.803  &  0.740 &  0.101 &   2.033 &  1.610 &  0.274 &	3.304  &2.420 &  0.482  & 4.366  &  3.190 &  0.724 &  4.439   &  3.934 &  0.997  \\ 
G2I &  1.050  &  0.825 &  0.090 &   2.295 &  1.682 &  0.265 &	3.566  &2.483 &  0.475  & 4.543  &  3.246 &  0.718 &  4.439   &  3.986 &  0.993  \\ 
G5I &  1.327  &  0.945 &  0.114 &   2.588 &  1.785 &  0.293 &	3.851  &2.575 &  0.507  & 4.711  &  3.331 &  0.754 &  4.425   &  4.067 &  1.032  \\ 
G8I &  1.760  &  1.108 &  0.183 &   3.016 &  1.948 &  0.371 &	4.225  &2.738 &  0.593  & 4.808  &  3.496 &  0.848 &  4.248   &  4.234 &  1.133  \\ 
K2I &  2.357  &  1.342 &  0.191 &   3.631 &  2.153 &  0.384 &	4.725  &2.922 &  0.611  & 4.871  &  3.665 &  0.870 &  4.070   &  4.392 &  1.159  \\ 
K3I &  2.451  &  1.442 &  0.243 &   3.719 &  2.255 &  0.442 &	4.759  &3.025 &  0.675  & 4.782  &  3.769 &  0.939 &  3.938   &  4.498 &  1.234  \\ 
K4I &  2.491  &  1.524 &  0.276 &   3.774 &  2.329 &  0.479 &	4.779  &3.095 &  0.715  & 4.707  &  3.837 &  0.983 &  3.830   &  4.565 &  1.281  \\ 
M2I &  2.641  &  1.624 &  0.483 &   3.876 &  2.433 &  0.703 &	4.701  &3.205 &  0.956  & 4.383  &  3.954 &  1.240 &  3.444   &  4.690 &  1.553  \\  \hline    
\end{tabular} 
\end{table*}      

\newpage

\begin{table*}
\caption[]{\UVEX/\IPHAS colour indices $(U-g)$, $(g-r)$, $(He${\sc i}$-r)$ $(r-H\alpha)$ and $(r-i)$ for log(g)=8.0 Bergeron DA white dwarfs including reddening.}
\begin{tabular}{p{3mm}p{9mm}p{9mm}p{9mm}p{9mm}p{9mm}p{7mm}p{9mm}p{9mm}p{9mm}p{9mm}p{7mm}p{7mm}p{9mm}p{7mm}p{9mm}}
\hline 
\multicolumn{1}{l}{T (K)}&
\multicolumn{5}{l}{E(B--V)=0.0}&
\multicolumn{5}{l}{E(B--V)=1.0}&
\multicolumn{5}{l}{E(B--V)=2.0}\\
   & $U$$-$$g$ & $g$$-$$r$ & $He${\sc i}$-$$r$ & $r$$-$$H\alpha$ & $r$$-$$i$ & $U$$-$$g$ & $g$$-$$r$ &
$He${\sc i}$-$$r$ & $r$$-$$H\alpha$ & $r$$-$$i$ &$U$$-$$g$ & $g$$-$$r$
& $He${\sc i}$-$$r$ & $r$$-$$H\alpha$
& $r$$-$$i$ \\ \hline
 1500 &  1.794 &  0.761 & --0.232 & --0.275 & --2.272  &  3.079 &  1.469 & --0.132 & --0.030 & --1.598 & 4.411  &  2.129 & --0.004 &  0.186 & --0.901 \\
 1750 &  1.581 &  1.025 & --0.095 &  0.007 & --1.613  &   2.865 &  1.772 &  0.036 &  0.220 & --0.944  &  4.190 &  2.469 &  0.196 &  0.404 & --0.254 \\
 2000 &  1.411 &  1.160 &  0.029 &  0.215 & --1.036  &    2.693 &  1.939 &  0.185 &  0.403 & --0.375  &  4.005 &  2.669 &  0.372 &  0.561 &  0.303 \\
 2250 &  1.267 &  1.229 &  0.118 &  0.325 & --0.531  &    2.545 &  2.034 &  0.293 &  0.495 &  0.124  &   3.843 &  2.788 &  0.499 &  0.634 &  0.792 \\
 2500 &  1.139 &  1.254 &  0.174 &  0.375 & --0.093  &    2.414 &  2.078 &  0.361 &  0.534 &  0.562  &   3.699 &  2.849 &  0.580 &  0.660 &  1.227 \\
 2750 &  1.031 &  1.249 &  0.203 &  0.393 &  0.258  &	  2.303 &  2.084 &  0.396 &  0.545 &  0.920  &   3.579 &  2.867 &  0.622 &  0.664 &  1.590 \\
 3000 &  0.939 &  1.225 &  0.214 &  0.396 &  0.501  &	  2.209 &  2.068 &  0.409 &  0.546 &  1.174  &   3.481 &  2.858 &  0.638 &  0.662 &  1.853 \\
 3250 &  0.855 &  1.191 &  0.214 &  0.393 &  0.638  &	  2.123 &  2.040 &  0.410 &  0.542 &  1.320  &   3.397 &  2.834 &  0.639 &  0.658 &  2.007 \\
 3500 &  0.774 &  1.151 &  0.209 &  0.386 &  0.694  &	  2.041 &  2.003 &  0.404 &  0.536 &  1.381  &   3.320 &  2.800 &  0.633 &  0.653 &  2.073 \\
 3750 &  0.695 &  1.106 &  0.201 &  0.378 &  0.700  &	  1.960 &  1.962 &  0.395 &  0.529 &  1.390  &   3.243 &  2.762 &  0.622 &  0.647 &  2.085 \\
 4000 &  0.612 &  1.057 &  0.191 &  0.368 &  0.682  &	  1.876 &  1.916 &  0.383 &  0.522 &  1.373  &   3.164 &  2.718 &  0.609 &  0.641 &  2.070 \\
 4250 &  0.523 &  1.002 &  0.180 &  0.357 &  0.650  &	  1.784 &  1.865 &  0.370 &  0.513 &  1.342  &   3.077 &  2.670 &  0.594 &  0.634 &  2.039 \\
 4500 &  0.425 &  0.941 &  0.168 &  0.344 &  0.610  &	  1.684 &  1.808 &  0.355 &  0.502 &  1.303  &   2.980 &  2.615 &  0.577 &  0.626 &  2.001 \\
 4750 &  0.322 &  0.876 &  0.155 &  0.329 &  0.567  &	  1.578 &  1.747 &  0.340 &  0.490 &  1.261  &   2.878 &  2.558 &  0.559 &  0.616 &  1.959 \\
 5000 &  0.218 &  0.810 &  0.143 &  0.311 &  0.523  &	  1.470 &  1.686 &  0.324 &  0.474 &  1.218  &   2.772 &  2.499 &  0.541 &  0.603 &  1.917 \\
 5250 &  0.118 &  0.747 &  0.130 &  0.288 &  0.482  &	  1.368 &  1.626 &  0.309 &  0.453 &  1.178  &   2.671 &  2.443 &  0.523 &  0.585 &  1.877 \\
 5500 &  0.027 &  0.688 &  0.118 &  0.260 &  0.445  &	  1.273 &  1.572 &  0.295 &  0.428 &  1.141  &   2.578 &  2.391 &  0.506 &  0.562 &  1.842 \\
 6000 & --0.136 &  0.584 &  0.098 &  0.205 &  0.381  &    1.105 &  1.474 &  0.270 &  0.377 &  1.079  &   2.410 &  2.298 &  0.477 &  0.516 &  1.780 \\
 6500 & --0.277 &  0.494 &  0.080 &  0.157 &  0.327  &    0.960 &  1.390 &  0.249 &  0.334 &  1.026  &   2.265 &  2.219 &  0.452 &  0.476 &  1.728 \\
 7000 & --0.381 &  0.420 &  0.066 &  0.117 &  0.280  &    0.853 &  1.321 &  0.232 &  0.297 &  0.980  &   2.157 &  2.152 &  0.432 &  0.442 &  1.684 \\
 7500 & --0.449 &  0.360 &  0.055 &  0.082 &  0.240  &    0.782 &  1.264 &  0.217 &  0.265 &  0.941  &   2.085 &  2.098 &  0.415 &  0.412 &  1.646 \\
 8000 & --0.489 &  0.310 &  0.044 &  0.048 &  0.204  &    0.740 &  1.216 &  0.204 &  0.233 &  0.906  &   2.042 &  2.052 &  0.400 &  0.383 &  1.611 \\
 8500 & --0.508 &  0.266 &  0.034 &  0.014 &  0.171  &    0.719 &  1.174 &  0.192 &  0.202 &  0.873  &   2.021 &  2.010 &  0.385 &  0.354 &  1.579 \\
 9000 & --0.511 &  0.228 &  0.023 & --0.022 &  0.140  &   0.715  &  1.136 &  0.178 &  0.167 &  0.843  &  2.016 &  1.972 &  0.369 &  0.322 &  1.550 \\
 9500 & --0.505 &  0.193 &  0.011 & --0.068 &  0.110  &   0.721  &  1.100 &  0.164 &  0.124 &  0.814  &  2.022 &  1.935 &  0.352 &  0.282 &  1.522 \\
10000 & --0.496 &  0.160 & --0.000 & --0.119 &  0.083  &  0.731  &  1.067 &  0.150 &  0.076 &  0.788  &  2.032 &  1.901 &  0.335 &  0.236 &  1.497\\
10500 & --0.489 &  0.132 & --0.011 & --0.161 &  0.057  &  0.737  &  1.037 &  0.137 &  0.037 &  0.764  &  2.039 &  1.869 &  0.320 &  0.199 &  1.474\\
11000 & --0.489 &  0.106 & --0.019 & --0.191 &  0.035  &  0.737  &  1.010 &  0.127 &  0.008 &  0.742  &  2.038 &  1.842 &  0.308 &  0.173 &  1.453\\
11500 & --0.496 &  0.082 & --0.026 & --0.212 &  0.016  &  0.730  &  0.987 &  0.119 & --0.011 &  0.724  & 2.032  &  1.818 &  0.298 &  0.155 &  1.435\\
12000 & --0.502 &  0.061 & --0.031 & --0.224 & --0.001  & 0.724  &  0.966 &  0.112 & --0.021 &  0.708  & 2.025  &  1.797 &  0.290 &  0.146 &  1.420\\
12500 & --0.509 &  0.044 & --0.034 & --0.225 & --0.013  & 0.716  &  0.949 &  0.108 & --0.022 &  0.696  & 2.017  &  1.781 &  0.286 &  0.146 &  1.408\\
13000 & --0.520 &  0.027 & --0.036 & --0.223 & --0.024  & 0.704  &  0.933 &  0.105 & --0.020 &  0.685  & 2.004  &  1.766 &  0.282 &  0.149 &  1.398\\
13500 & --0.538 &  0.010 & --0.038 & --0.221 & --0.034  & 0.685  &  0.918 &  0.103 & --0.017 &  0.675  & 1.985  &  1.752 &  0.279 &  0.153 &  1.387\\
14000 & --0.562 & --0.005 & --0.039 & --0.217 & --0.044  & 0.660  &  0.905 &  0.102 & --0.013 &  0.665  & 1.958  &  1.740 &  0.278 &  0.157 &  1.378\\
14500 & --0.589 & --0.019 & --0.039 & --0.212 & --0.051  & 0.631  &  0.893 &  0.102 & --0.007 &  0.657  & 1.929  &  1.730 &  0.277 &  0.163 &  1.370\\
15000 & --0.618 & --0.031 & --0.039 & --0.206 & --0.058  & 0.601  &  0.882 &  0.102 & --0.001 &  0.650  & 1.898  &  1.721 &  0.277 &  0.169 &  1.363\\
15500 & --0.647 & --0.042 & --0.038 & --0.199 & --0.064  & 0.571  &  0.873 &  0.102 &  0.006 &  0.644  & 1.866  &  1.713 &  0.278 &  0.176 &  1.356\\
16000 & --0.676 & --0.052 & --0.038 & --0.192 & --0.070  & 0.541  &  0.865 &  0.102 &  0.013 &  0.638  & 1.836  &  1.706 &  0.278 &  0.183 &  1.350\\
16500 & --0.704 & --0.061 & --0.038 & --0.184 & --0.075  & 0.512  &  0.857 &  0.103 &  0.020 &  0.633  & 1.806  &  1.700 &  0.278 &  0.190 &  1.345\\
17000 & --0.732 & --0.070 & --0.038 & --0.177 & --0.079  & 0.483  &  0.850 &  0.103 &  0.027 &  0.629  & 1.777  &  1.695 &  0.279 &  0.197&  1.340  \\ \hline  
\end{tabular} 
\end{table*}

\newpage

\addtocounter{table}{-1}
\begin{table*}
\caption[]{, continued}
\flushleft
\begin{tabular}{ccccccccccc}
\hline 
\multicolumn{1}{l}{T (K)}&
\multicolumn{5}{l}{E(B--V)=3.0}&
\multicolumn{5}{l}{E(B--V)=4.0}\\
      & $U$$-$$g$ & $g$$-$$r$ & $He${\sc i}$-$$r$ & $r$$-$$H\alpha$ &
$r$$-$$i$ & $U$$-$$g$ & $g$$-$$r$ & $He${\sc i}$-$$r$ & $r$$-$$H\alpha$ & $r$$-$$i$\\ \hline
 1500 &  5.679  &  2.761 &  0.153 &  0.374 & -0.179  &   6.352 &  3.380 &  0.339 &  0.532 &  0.566 \\	     
 1750 &  5.381  &  3.136 &  0.385 &  0.559 &  0.455  &   5.766 &  3.789 &  0.604 &  0.685 &  1.184 \\	     
 2000 &  5.108  &  3.368 &  0.589 &  0.689 &  0.998  &   5.253 &  4.050 &  0.834 &  0.788 &  1.708 \\	     
 2250 &  4.854  &  3.511 &  0.736 &  0.742 &  1.474  &   4.807 &  4.216 &  1.002 &  0.821 &  2.169 \\	     
 2500 &  4.630  &  3.589 &  0.830 &  0.755 &  1.903  &   4.448 &  4.310 &  1.110 &  0.821 &  2.590 \\	     
 2750 &  4.460  &  3.617 &  0.880 &  0.752 &  2.269  &   4.203 &  4.348 &  1.166 &  0.811 &  2.955 \\	     
 3000 &  4.347  &  3.614 &  0.899 &  0.747 &  2.538  &   4.067 &  4.350 &  1.189 &  0.802 &  3.229 \\	     
 3250 &  4.277  &  3.593 &  0.901 &  0.742 &  2.699  &   4.012 &  4.332 &  1.191 &  0.797 &  3.396 \\	     
 3500 &  4.230  &  3.562 &  0.893 &  0.737 &  2.770  &   4.006 &  4.303 &  1.184 &  0.793 &  3.471 \\	     
 3750 &  4.192  &  3.525 &  0.881 &  0.733 &  2.784  &   4.024 &  4.267 &  1.171 &  0.789 &  3.488 \\	     
 4000 &  4.154  &  3.483 &  0.867 &  0.729 &  2.770  &   4.054 &  4.226 &  1.154 &  0.786 &  3.475 \\	     
 4250 &  4.111  &  3.436 &  0.850 &  0.723 &  2.741  &   4.089 &  4.180 &  1.136 &  0.783 &  3.446 \\	     
 4500 &  4.058  &  3.383 &  0.830 &  0.717 &  2.703  &   4.126 &  4.128 &  1.114 &  0.779 &  3.409 \\	     
 4750 &  3.996  &  3.327 &  0.810 &  0.710 &  2.661  &   4.160 &  4.073 &  1.092 &  0.773 &  3.368 \\	     
 5000 &  3.927  &  3.271 &  0.790 &  0.699 &  2.620  &   4.189 &  4.017 &  1.069 &  0.765 &  3.327 \\	     
 5250 &  3.856  &  3.216 &  0.770 &  0.684 &  2.581  &   4.210 &  3.964 &  1.047 &  0.752 &  3.289 \\	     
 5500 &  3.786  &  3.166 &  0.751 &  0.663 &  2.546  &   4.223 &  3.914 &  1.026 &  0.733 &  3.254 \\	     
 6000 &  3.654  &  3.076 &  0.718 &  0.621 &  2.486  &   4.233 &  3.826 &  0.989 &  0.694 &  3.195 \\	     
 6500 &  3.533  &  2.999 &  0.689 &  0.584 &  2.435  &   4.225 &  3.750 &  0.958 &  0.661 &  3.145 \\	     
 7000 &  3.441  &  2.935 &  0.666 &  0.553 &  2.392  &   4.215 &  3.687 &  0.932 &  0.633 &  3.103 \\	     
 7500 &  3.378  &  2.882 &  0.647 &  0.526 &  2.354  &   4.210 &  3.635 &  0.911 &  0.608 &  3.066 \\	     
 8000 &  3.341  &  2.837 &  0.629 &  0.499 &  2.320  &   4.213 &  3.589 &  0.891 &  0.583 &  3.032 \\	     
 8500 &  3.324  &  2.795 &  0.612 &  0.473 &  2.289  &   4.224 &  3.548 &  0.872 &  0.559 &  3.002 \\	     
 9000 &  3.323  &  2.756 &  0.594 &  0.443 &  2.260  &   4.243 &  3.507 &  0.851 &  0.531 &  2.974 \\	     
 9500 &  3.331  &  2.718 &  0.574 &  0.405 &  2.234  &   4.266 &  3.468 &  0.830 &  0.495 &  2.948 \\	     
10000 &  3.343   &  2.682 &  0.555 &  0.362 &  2.209  &  4.291  &  3.430 &  0.809 &  0.455 &  2.925 \\       
10500 &  3.352   &  2.649 &  0.538 &  0.327 &  2.187  &  4.312  &  3.396 &  0.789 &  0.422 &  2.903 \\       
11000 &  3.354   &  2.621 &  0.524 &  0.302 &  2.167  &  4.326  &  3.366 &  0.773 &  0.399 &  2.884 \\       
11500 &  3.349   &  2.596 &  0.512 &  0.287 &  2.150  &  4.334  &  3.341 &  0.760 &  0.384 &  2.867 \\       
12000 &  3.344   &  2.575 &  0.503 &  0.279 &  2.135  &  4.340  &  3.319 &  0.750 &  0.378 &  2.853 \\       
12500 &  3.337   &  2.559 &  0.498 &  0.280 &  2.124  &  4.342  &  3.303 &  0.744 &  0.379 &  2.842 \\       
13000 &  3.325   &  2.544 &  0.493 &  0.283 &  2.114  &  4.340  &  3.288 &  0.739 &  0.383 &  2.832 \\       
13500 &  3.306   &  2.531 &  0.490 &  0.287 &  2.103  &  4.332  &  3.276 &  0.735 &  0.388 &  2.822 \\       
14000 &  3.281   &  2.520 &  0.489 &  0.292 &  2.094  &  4.319  &  3.266 &  0.733 &  0.393 &  2.813 \\       
14500 &  3.252   &  2.511 &  0.488 &  0.298 &  2.086  &  4.302  &  3.258 &  0.733 &  0.399 &  2.804 \\       
15000 &  3.222   &  2.504 &  0.488 &  0.304 &  2.078  &  4.283  &  3.252 &  0.732 &  0.405 &  2.797 \\       
15500 &  3.192   &  2.498 &  0.488 &  0.311 &  2.072  &  4.263  &  3.246 &  0.733 &  0.413 &  2.790 \\       
16000 &  3.162   &  2.492 &  0.488 &  0.318 &  2.066  &  4.243  &  3.242 &  0.733 &  0.420 &  2.784 \\       
16500 &  3.133   &  2.487 &  0.489 &  0.326 &  2.060  &  4.224  &  3.237 &  0.733 &  0.427 &  2.778 \\       
17000 &  3.105   &  2.482 &  0.489 &  0.333 &  2.055  &  4.204  &  3.233 &  0.733 &  0.434 &  2.773 \\ \hline
\end{tabular}
\end{table*}

\newpage

\begin{table*}
\caption[]{\UVEX/\IPHAS colour indices $(U-g)$, $(g-r)$, $(HeI-r)$
  $(r-H\alpha)$ and $(r-i)$ for log(g)=8.0 Koester DA white dwarfs
  including reddening}
\begin{tabular}{p{3mm}p{9mm}p{9mm}p{9mm}p{9mm}p{9mm}p{9mm}p{7mm}p{9mm}p{9mm}p{7mm}p{7mm}p{7mm}p{9mm}p{7mm}p{7mm}}
\hline 
\multicolumn{1}{l}{T (K)}&
\multicolumn{5}{l}{E(B--V)=0.0}&
\multicolumn{5}{l}{E(B--V)=1.0}&
\multicolumn{5}{l}{E(B--V)=2.0}\\
 & $U$$-$$g$ & $g$$-$$r$ & $He${\sc i}$-$$r$ & $r$$-$$H\alpha$ & $r$$-$$i$ & $U$$-$$g$ & $g$$-$$r$
 & $He${\sc i}$-$$r$ & $r$$-$$H\alpha$ & $r$$-$$i$ & $U$$-$$g$ &
$g$$-$$r$ & $He${\sc i}$-$$r$ &
 $r$$-$$H\alpha$ & $r$$-$$i$ \\\hline
 6000 & --0.118 &  0.591 &  0.102 &  0.257 &  0.378  &        1.123  &  1.482 &  0.275 &  0.429 &  1.074  &   2.428 &  2.306 &  0.483 &  0.566 &  1.775 \\
 7000 & --0.375 &  0.423 &  0.072 &  0.191 &  0.275  &        0.858  &  1.325 &  0.238 &  0.369 &  0.974  &   2.161 &  2.158 &  0.439 &  0.513 &  1.677 \\
 8000 & --0.487 &  0.312 &  0.047 &  0.101 &  0.200  &        0.742  &  1.219 &  0.208 &  0.286 &  0.901  &   2.044 &  2.056 &  0.404 &  0.435 &  1.606 \\
 9000 & --0.510 &  0.228 &  0.024 & --0.009 &  0.139  &       0.717  &  1.136 &  0.179 &  0.180 &  0.841  &   2.018 &  1.973 &  0.370 &  0.335 &  1.548 \\
10000 & --0.492 &  0.160 &  0.000 & --0.117 &  0.083  &       0.735  &  1.066 &  0.151 &  0.078 &  0.788  &   2.036 &  1.900 &  0.336 &  0.239 &  1.497 \\
11000 & --0.487 &  0.105 & --0.018 & --0.183 &  0.035  &      0.739  &  1.010 &  0.129 &  0.017 &  0.742  &   2.041 &  1.841 &  0.310 &  0.181 &  1.453 \\
12000 & --0.504 &  0.063 & --0.028 & --0.207 &  0.002  &      0.722  &  0.968 &  0.116 & --0.005 &  0.710  &  2.024 &  1.799 &  0.294 &  0.162 &  1.422 \\
13000 & --0.527 &  0.028 & --0.033 & --0.208 & --0.022  &     0.698  &  0.935 &  0.109 & --0.005 &  0.687  &  1.998 &  1.768 &  0.286 &  0.164 &  1.400 \\
14000 & --0.567 & --0.004 & --0.037 & --0.205 & --0.042  &    0.655  &  0.905 &  0.104 & --0.001 &  0.667  &  1.954 &  1.741 &  0.281 &  0.168 &  1.379 \\
15000 & --0.622 & --0.030 & --0.037 & --0.196 & --0.057  &    0.598  &  0.883 &  0.104 &  0.008 &  0.651  &   1.895 &  1.722 &  0.279 &  0.178 &  1.364 \\
16000 & --0.681 & --0.051 & --0.037 & --0.184 & --0.069  &    0.537  &  0.866 &  0.104 &  0.021 &  0.639  &   1.833 &  1.707 &  0.280 &  0.191 &  1.351 \\
17000 & --0.738 & --0.069 & --0.036 & --0.170 & --0.078  &    0.478  &  0.851 &  0.104 &  0.034 &  0.629  &   1.773 &  1.695 &  0.280 &  0.204 &  1.341 \\
18000 & --0.791 & --0.084 & --0.036 & --0.158 & --0.087  &    0.423  &  0.838 &  0.105 &  0.047 &  0.621  &   1.717 &  1.685 &  0.280 &  0.217 &  1.332 \\
19000 & --0.840 & --0.098 & --0.036 & --0.147 & --0.094  &    0.373  &  0.827 &  0.105 &  0.057 &  0.613  &   1.666 &  1.675 &  0.280 &  0.227 &  1.324 \\
20000 & --0.885 & --0.110 & --0.036 & --0.138 & --0.101  &    0.327  &  0.816 &  0.105 &  0.067 &  0.606  &   1.619 &  1.666 &  0.280 &  0.237 &  1.317 \\
22000 & --0.964 & --0.132 & --0.036 & --0.124 & --0.114  &    0.246  &  0.798 &  0.104 &  0.081 &  0.593  &   1.537 &  1.651 &  0.279 &  0.251 &  1.304 \\
24000 & --1.031 & --0.151 & --0.037 & --0.114 & --0.125  &    0.178  &  0.781 &  0.103 &  0.091 &  0.581  &   1.467 &  1.637 &  0.278 &  0.262 &  1.292 \\
26000 & --1.091 & --0.170 & --0.037 & --0.106 & --0.136  &    0.116  &  0.766 &  0.103 &  0.100 &  0.571  &   1.405 &  1.623 &  0.277 &  0.270 &  1.281 \\
28000 & --1.146 & --0.185 & --0.037 & --0.095 & --0.145  &    0.059  &  0.752 &  0.102 &  0.111 &  0.561  &   1.347 &  1.612 &  0.277 &  0.282 &  1.271 \\
30000 & --1.197 & --0.198 & --0.037 & --0.078 & --0.152  &    0.008  &  0.742 &  0.103 &  0.127 &  0.553  &   1.294 &  1.603 &  0.277 &  0.298 &  1.263 \\
35000 & --1.278 & --0.219 & --0.035 & --0.038 & --0.163  &  --0.077 &  0.725 &  0.105 &  0.167 &  0.542  &    1.208 &  1.590 &  0.279 &  0.337 &  1.252 \\
40000 & --1.318 & --0.231 & --0.034 & --0.017 & --0.169  &  --0.118 &  0.716 &  0.106 &  0.188 &  0.536  &    1.166 &  1.583 &  0.280 &  0.359 &  1.245 \\
45000 & --1.342 & --0.239 & --0.034 & --0.004 & --0.172  &  --0.143 &  0.709 &  0.106 &  0.201 &  0.532  &    1.140 &  1.578 &  0.281 &  0.372 &  1.241 \\
50000 & --1.359 & --0.244 & --0.034 &  0.005 & --0.175  &   --0.160 &  0.704 &  0.106 &  0.210 &  0.529  &    1.122 &  1.574 &  0.281 &  0.381 &  1.238 \\
55000 & --1.372 & --0.249 & --0.034 &  0.012 & --0.177  &   --0.174 &  0.700 &  0.107 &  0.217 &  0.527  &    1.108 &  1.570 &  0.281 &  0.387 &  1.236 \\
60000 & --1.382 & --0.253 & --0.033 &  0.017 & --0.179  &   --0.185 &  0.697 &  0.107 &  0.222 &  0.525  &    1.097 &  1.568 &  0.281 &  0.392 &  1.234 \\
65000 & --1.391 & --0.256 & --0.033 &  0.021 & --0.181  &   --0.193 &  0.694 &  0.107 &  0.226 &  0.523  &    1.088 &  1.565 &  0.282 &  0.397 &  1.232 \\
70000 & --1.398 & --0.259 & --0.033 &  0.025 & --0.183  &   --0.201 &  0.692 &  0.107 &  0.230 &  0.522  &    1.080 &  1.563 &  0.282 &  0.400 &  1.230 \\
75000 & --1.404 & --0.262 & --0.033 &  0.027 & --0.184  &   --0.208 &  0.690 &  0.107 &  0.232 &  0.520  &    1.073 &  1.562 &  0.281 &  0.403 &  1.229 \\
80000 & --1.410 & --0.264 & --0.033 &  0.030 & --0.185  &   --0.213 &  0.688 &  0.107 &  0.235 &  0.519  &    1.067 &  1.560 &  0.281 &  0.405 &  1.227 \\ \hline    
\end{tabular}
\end{table*}

\newpage

\addtocounter{table}{-1}
\begin{table*}
\caption[]{, continued}
\flushleft
\begin{tabular}{cccccccccccccccc}
\hline 
\multicolumn{1}{l}{T (K)}&
\multicolumn{5}{l}{E(B--V)=3.0}&
\multicolumn{5}{l}{E(B--V)=4.0}\\
      & $U$$-$$g$ & $g$$-$$r$ & $He${\sc i}$-$$r$ & $r$$-$$H\alpha$ &
$r$$-$$i$ &$U$$-$$g$ & $g$$-$$r$ & $He${\sc i}$-$$r$ & $r$$-$$H\alpha$ & $r$$-$$i$\\ \hline
 6000 &   3.669 &  3.084 &  0.724 &  0.671 &  2.480  &   4.238  &  3.834 &  0.996 &  0.744 &  3.189 \\
 7000 &   3.444 &  2.942 &  0.674 &  0.624 &  2.384  &   4.217  &  3.695 &  0.941 &  0.703 &  3.094 \\
 8000 &   3.343 &  2.841 &  0.634 &  0.551 &  2.314  &   4.214  &  3.594 &  0.896 &  0.634 &  3.026 \\
 9000 &   3.325 &  2.757 &  0.595 &  0.456 &  2.259  &   4.244  &  3.508 &  0.853 &  0.544 &  2.972 \\
10000 &   3.348 &  2.681 &  0.556 &  0.364 &  2.210  &   4.294  &  3.429 &  0.809 &  0.457 &  2.925\\
11000 &   3.357 &  2.620 &  0.526 &  0.311 &  2.167  &   4.328  &  3.366 &  0.775 &  0.407 &  2.884\\
12000 &   3.343 &  2.577 &  0.507 &  0.294 &  2.137  &   4.338  &  3.322 &  0.755 &  0.393 &  2.855\\
13000 &   3.319 &  2.547 &  0.498 &  0.297 &  2.115  &   4.335  &  3.292 &  0.744 &  0.397 &  2.834\\
14000 &   3.278 &  2.521 &  0.492 &  0.303 &  2.095  &   4.316  &  3.268 &  0.737 &  0.404 &  2.814\\
15000 &   3.220 &  2.505 &  0.490 &  0.313 &  2.079  &   4.282  &  3.253 &  0.735 &  0.414 &  2.798\\
16000 &   3.160 &  2.493 &  0.490 &  0.326 &  2.067  &   4.242  &  3.243 &  0.735 &  0.427 &  2.785\\
17000 &   3.101 &  2.483 &  0.490 &  0.339 &  2.056  &   4.202  &  3.234 &  0.735 &  0.441 &  2.774\\
18000 &   3.047 &  2.474 &  0.491 &  0.352 &  2.047  &   4.164  &  3.227 &  0.735 &  0.453 &  2.765\\
19000 &   2.997 &  2.466 &  0.491 &  0.363 &  2.039  &   4.128  &  3.220 &  0.735 &  0.464 &  2.756\\
20000 &   2.951 &  2.459 &  0.490 &  0.372 &  2.031  &   4.095  &  3.213 &  0.734 &  0.474 &  2.749\\
22000 &   2.871 &  2.445 &  0.489 &  0.386 &  2.018  &   4.035  &  3.202 &  0.733 &  0.488 &  2.735\\
24000 &   2.803 &  2.433 &  0.488 &  0.397 &  2.006  &   3.983  &  3.191 &  0.731 &  0.500 &  2.723\\
26000 &   2.741 &  2.421 &  0.487 &  0.407 &  1.995  &   3.936  &  3.180 &  0.730 &  0.509 &  2.712\\
28000 &   2.684 &  2.411 &  0.486 &  0.418 &  1.985  &   3.892  &  3.171 &  0.729 &  0.521 &  2.702\\
30000 &   2.632 &  2.404 &  0.486 &  0.434 &  1.977  &   3.849  &  3.166 &  0.729 &  0.537 &  2.693\\
35000 &   2.546 &  2.395 &  0.489 &  0.473 &  1.964  &   3.778  &  3.158 &  0.732 &  0.576 &  2.681\\
40000 &   2.503 &  2.389 &  0.490 &  0.495 &  1.958  &   3.743  &  3.154 &  0.733 &  0.597 &  2.674\\
45000 &   2.478 &  2.384 &  0.490 &  0.508 &  1.953  &   3.721  &  3.150 &  0.733 &  0.610 &  2.669\\
50000 &   2.459 &  2.381 &  0.491 &  0.516 &  1.950  &   3.706  &  3.148 &  0.734 &  0.619 &  2.666\\
55000 &   2.446 &  2.379 &  0.491 &  0.523 &  1.948  &   3.694  &  3.145 &  0.734 &  0.625 &  2.663\\
60000 &   2.434 &  2.376 &  0.491 &  0.528 &  1.946  &   3.685  &  3.144 &  0.734 &  0.631 &  2.661\\
65000 &   2.425 &  2.375 &  0.491 &  0.532 &  1.944  &   3.677  &  3.142 &  0.734 &  0.635 &  2.660\\
70000 &   2.417 &  2.373 &  0.491 &  0.536 &  1.942  &   3.670  &  3.141 &  0.734 &  0.638 &  2.658\\
75000 &   2.410 &  2.371 &  0.491 &  0.539 &  1.941  &   3.665  &  3.139 &  0.734 &  0.641 &  2.656\\
80000 &   2.404 &  2.370 &  0.491 &  0.541 &  1.940  &   3.659  &  3.138 &  0.734 &  0.643 &  2.655\\ \hline
\end{tabular}
\end{table*}

\clearpage
\newpage

\begin{table*}
\caption[]{\UVEX/\IPHAS colour indices $(U-g)$, $(g-r)$, $(HeI-r)$ $(r-H\alpha)$ and $(r-i)$ for log(g)=8.0 Koester DB white dwarfs including reddening.}
\begin{tabular}{p{3mm}p{9mm}p{9mm}p{9mm}p{7mm}p{9mm}p{9mm}p{7mm}p{9mm}p{7mm}p{7mm}p{7mm}p{7mm}p{9mm}p{7mm}p{7mm}}
\hline 
\multicolumn{1}{l}{T (K)}&
\multicolumn{5}{l}{E(B--V)=0.0}&
\multicolumn{5}{l}{E(B--V)=1.0}&
\multicolumn{5}{l}{E(B--V)=2.0}\\
      & $U$$-$$g$ & $g$$-$$r$ & $He${\sc i}$-$$r$ & $r$$-$$H\alpha$ & $r$$-$$i$ & $U$$-$$g$ & $g$$-$$r$ &
$He${\sc i}$-$$r$ & $r$$-$$H\alpha$ & $r$$-$$i$ & $U$$-$$g$ &
$g$$-$$r$ & $He${\sc i}$-$$r$ & $r$$-$$H\alpha$
& $r$$-$$i$ \\ \hline
10000 & --0.753 &  0.162 &  0.033 &  0.172 &  0.081  &      0.465 &  1.083 &  0.189 &  0.361 &  0.781  &   1.759 &  1.930 &  0.380 &  0.515 &  1.485  \\
11000 & --0.839 &  0.109 &  0.031 &  0.161 &  0.049  &      0.376 &  1.033 &  0.186 &  0.351 &  0.749  &   1.669 &  1.883 &  0.375 &  0.507 &  1.453  \\
12000 & --0.907 &  0.067 &  0.037 &  0.153 &  0.023  &      0.305 &  0.995 &  0.190 &  0.345 &  0.724  &   1.597 &  1.848 &  0.378 &  0.502 &  1.428  \\
13000 & --0.955 &  0.032 &  0.050 &  0.146 & --0.000  &     0.256 &  0.962 &  0.202 &  0.339 &  0.701  &   1.547 &  1.816 &  0.388 &  0.498 &  1.406  \\
14000 & --0.990 &  0.001 &  0.072 &  0.140 & --0.020  &     0.219 &  0.933 &  0.223 &  0.334 &  0.681  &   1.510 &  1.788 &  0.408 &  0.494 &  1.386  \\
15000 & --1.014 & --0.024 &  0.101 &  0.135 & --0.037  &    0.195 &  0.908 &  0.250 &  0.330 &  0.665  &   1.485 &  1.764 &  0.435 &  0.491 &  1.370  \\
16000 & --1.028 & --0.045 &  0.134 &  0.130 & --0.050  &    0.180 &  0.888 &  0.283 &  0.326 &  0.652  &   1.471 &  1.744 &  0.466 &  0.488 &  1.358  \\
17000 & --1.037 & --0.063 &  0.167 &  0.125 & --0.060  &    0.172 &  0.870 &  0.314 &  0.322 &  0.643  &   1.463 &  1.725 &  0.497 &  0.485 &  1.349  \\
18000 & --1.043 & --0.078 &  0.194 &  0.120 & --0.068  &    0.166 &  0.855 &  0.341 &  0.318 &  0.635  &   1.458 &  1.709 &  0.522 &  0.481 &  1.343  \\
19000 & --1.048 & --0.091 &  0.212 &  0.115 & --0.075  &    0.161 &  0.841 &  0.358 &  0.314 &  0.629  &   1.453 &  1.695 &  0.539 &  0.478 &  1.337  \\
20000 & --1.054 & --0.103 &  0.220 &  0.111 & --0.081  &    0.155 &  0.830 &  0.366 &  0.310 &  0.624  &   1.447 &  1.684 &  0.546 &  0.475 &  1.332  \\
22000 & --1.064 & --0.117 &  0.216 &  0.107 & --0.088  &    0.144 &  0.816 &  0.361 &  0.306 &  0.617  &   1.436 &  1.670 &  0.541 &  0.471 &  1.325  \\
24000 & --1.077 & --0.128 &  0.199 &  0.104 & --0.094  &    0.131 &  0.807 &  0.343 &  0.304 &  0.611  &   1.422 &  1.662 &  0.523 &  0.469 &  1.319  \\
26000 & --1.100 & --0.142 &  0.185 &  0.100 & --0.103  &    0.107 &  0.793 &  0.329 &  0.301 &  0.602  &   1.397 &  1.650 &  0.508 &  0.467 &  1.310  \\
28000 & --1.127 & --0.156 &  0.175 &  0.098 & --0.113  &    0.078 &  0.781 &  0.319 &  0.299 &  0.592  &   1.368 &  1.639 &  0.497 &  0.465 &  1.300  \\
30000 & --1.156 & --0.169 &  0.164 &  0.096 & --0.122  &    0.048 &  0.770 &  0.307 &  0.297 &  0.583  &   1.336 &  1.630 &  0.485 &  0.464 &  1.291  \\
35000 & --1.227 & --0.196 &  0.134 &  0.092 & --0.139  &   --0.025 &  0.747 &  0.277 &  0.294 &  0.565  &  1.261 &  1.611 &  0.454 &  0.461 &  1.274  \\
40000 & --1.285 & --0.218 &  0.116 &  0.088 & --0.154  &   --0.085 &  0.728 &  0.258 &  0.290 &  0.551  &  1.199 &  1.594 &  0.435 &  0.459 &  1.259  \\
50000 & --1.418 & --0.236 &  0.038 &  0.080 & --0.180  &   --0.223 &  0.717 &  0.180 &  0.284 &  0.524  &  1.058 &  1.589 &  0.355 &  0.453 &  1.232  \\ \hline    
\end{tabular} 
\end{table*}

\newpage

\addtocounter{table}{-1}
\begin{table*}
\caption[]{, continued}
\flushleft
\begin{tabular}{ccccccccccc} \hline
\multicolumn{1}{l}{T (K)}&
\multicolumn{5}{l}{E(B--V)=3.0}&
\multicolumn{5}{l}{E(B--V)=4.0}\\
 &$U$$-$$g$ & $g$$-$$r$ & $He${\sc i}$-$$r$ & $r$$-$$H\alpha$ &
$r$$-$$i$ & $U$$-$$g$ & $g$$-$$r$ & $He${\sc i}$-$$r$ & $r$$-$$H\alpha$ & $r$$-$$i$\\ \hline
10000 &  3.076 &  2.722 &  0.605 &  0.635 &  2.193  &	4.099 &  3.481 &  0.863 &  0.723 &  2.904\\
11000 &  2.991 &  2.678 &  0.598 &  0.629 &  2.162  &	4.055 &  3.438 &  0.854 &  0.719 &  2.874\\
12000 &  2.922 &  2.645 &  0.600 &  0.626 &  2.137  &	4.016 &  3.406 &  0.854 &  0.716 &  2.849\\
13000 &  2.875 &  2.614 &  0.609 &  0.622 &  2.114  &	3.991 &  3.375 &  0.863 &  0.714 &  2.827\\
14000 &  2.839 &  2.587 &  0.627 &  0.620 &  2.095  &	3.972 &  3.348 &  0.880 &  0.713 &  2.808\\
15000 &  2.816 &  2.563 &  0.653 &  0.618 &  2.080  &	3.961 &  3.324 &  0.904 &  0.712 &  2.793\\
16000 &  2.803 &  2.542 &  0.684 &  0.616 &  2.068  &	3.958 &  3.302 &  0.934 &  0.711 &  2.781\\
17000 &  2.797 &  2.523 &  0.713 &  0.613 &  2.060  &	3.959 &  3.282 &  0.963 &  0.709 &  2.774\\
18000 &  2.794 &  2.505 &  0.738 &  0.610 &  2.054  &	3.962 &  3.264 &  0.988 &  0.707 &  2.768\\
19000 &  2.790 &  2.491 &  0.754 &  0.607 &  2.048  &	3.964 &  3.248 &  1.003 &  0.704 &  2.763\\
20000 &  2.784 &  2.479 &  0.761 &  0.605 &  2.044  &	3.962 &  3.236 &  1.009 &  0.702 &  2.759\\
22000 &  2.774 &  2.465 &  0.755 &  0.602 &  2.037  &	3.958 &  3.222 &  1.002 &  0.699 &  2.752\\
24000 &  2.760 &  2.458 &  0.737 &  0.600 &  2.031  &	3.948 &  3.215 &  0.984 &  0.698 &  2.747\\
26000 &  2.736 &  2.447 &  0.722 &  0.598 &  2.022  &	3.930 &  3.205 &  0.969 &  0.696 &  2.738\\
28000 &  2.706 &  2.437 &  0.710 &  0.597 &  2.012  &	3.907 &  3.196 &  0.957 &  0.696 &  2.728\\
30000 &  2.675 &  2.429 &  0.698 &  0.596 &  2.003  &	3.883 &  3.189 &  0.944 &  0.695 &  2.719\\
35000 &  2.599 &  2.413 &  0.666 &  0.594 &  1.986  &	3.822 &  3.176 &  0.911 &  0.694 &  2.701\\
40000 &  2.538 &  2.398 &  0.646 &  0.592 &  1.971  &	3.772 &  3.162 &  0.890 &  0.693 &  2.686\\
50000 &  2.394 &  2.399 &  0.566 &  0.588 &  1.943  &	3.646 &  3.168 &  0.810 &  0.689 &  2.658\\ \hline
\end{tabular}
\end{table*}

\end{document}